\documentclass[3p]{elsarticle}
\usepackage{lineno,hyperref}
\hypersetup{
colorlinks = true,
linkcolor = blue,
anchorcolor = blue,
citecolor = blue,
filecolor = blue,
urlcolor = blue
}
\usepackage{xcolor}
\usepackage{amsmath}
\usepackage{amssymb}
\usepackage{float}
\usepackage{bm}
\usepackage{multirow}
\usepackage{subcaption}
\usepackage{cuted}
\usepackage[colorinlistoftodos]{todonotes}

\usepackage{algorithm}
\usepackage{algpseudocode}
\usepackage{comment}

\algdef{SE}[SWITCH]{Switch}{EndSwitch}[1]{\algorithmicswitch\ #1\ \algorithmicdo}{\algorithmicend\ \algorithmicswitch}%
\algdef{SE}[CASE]{Case}{EndCase}[1]{\algorithmiccase\ #1}{\algorithmicend\ \algorithmiccase}%
\algtext*{EndSwitch}%
\algtext*{EndCase}%

\usepackage[nameinlink,capitalize]{cleveref}
\usepackage{placeins}
\usepackage{pdfpages}
\usepackage{caption}
\usepackage[normalem]{ulem}
\useunder{\uline}{\ul}{}
\journal{Journal of the Mechanics and Physics of Solids}
\newcommand{\boldface}[1]{\boldsymbol{#1}}  % italic (slanted)
\newcommand{\bfa}{\boldface{a}}

\newcommand{\bfl}{\boldface{l}}
\newcommand{\bfm}{\boldface{m}}
\newcommand{\bfn}{\boldface{n}}

\newcommand{\bfp}{\boldface{p}}

\newcommand{\bfs}{\boldface{s}}

\newcommand{\bfC}{\boldface{C}}

\newcommand{\bfF}{\boldface{F}}

\newcommand{\bfP}{\boldface{P}}
\newcommand{\bfQ}{\boldface{Q}}
\newcommand{\bfR}{\boldface{R}}

\newcommand{\bfepsilon}{\boldsymbol{\epsilon}}

\newcommand{\bflambda}{\boldsymbol{\lambda}}

\newcommand{\bfxi}{\boldsymbol{\xi}}
\newcommand{\bfsigma}{\boldsymbol{\sigma}}

\newcommand{\bfSigma}{\boldsymbol{\Sigma}}

\newcommand{\T}{^{\mathrm{T}}} % x^{T}
\newlength{\boxwidth}
\def\dd{\;\!\mathrm{d}}

\def\btheorem{\begin{theorem}}
\def\etheorem{\end{theorem}}
\def\blemma{\begin{lemma}}
\def\elemma{\end{lemma}}
\def\bproposition{\begin{proposition}}
\def\eproposition{\end{proposition}}
\def\bcorollary{\begin{corollary}}
\def\ecorollary{\end{corollary}}
\def\bdefinition{\begin{definition}}
\def\edefinition{\end{definition}}
\def\bexample{\begin{example}}
\def\eexample{\end{example}}
\def\bremark{\begin{remark}}
\def\eremark{\end{remark}}

\DeclareMathOperator{\tr}{tr}

\newcommand{\be}{\begin{equation}}
\newcommand{\ee}{\end{equation}}
\newcommand{\beq}{\begin{eqnarray}}
\newcommand{\eeq}{\end{eqnarray}}
\newcommand{\bem}{\begin{multline}}
\newcommand{\eem}{\end{multline}}
\newcommand{\ba}{\begin{align}}
\newcommand{\ea}{\end{align}}

\newtheorem{theorem}{Theorem}
\newtheorem{definition}{Definition}

\newcommand{\st}[1]{\qquad\text{#1}\qquad}

\begin{document}
\modulolinenumbers[5]

\begin{frontmatter}

\title{Temperature-Aware Recurrent Neural Operator for Temperature-Dependent Anisotropic Plasticity in HCP Materials}

\author[1]{Yannick Hollenweger}
\author[1]{Dennis M. Kochmann}
\author[2]{Burigede Liu\corref{corrauthor}}

\cortext[corrauthor]{Corresponding author. E-mail address: bl377@cam.ac.uk}

\affiliation[1]{organization={Mechanics \& Materials Laboratory, Department of Mechanical and Process Engineering, ETH Zürich},
            city={Zürich},
            postcode={8092},
            country={Switzerland}}

\affiliation[2]{organization={Department of Engineering, University of Cambridge},
            addressline={Trumpington Street},
            city={Cambridge},
            postcode={CB2 1PZ},
            country={UK}}

\begin{abstract}
% Neural network surrogate models for constitutive models in computational mechanics have been developed and employed for a while. Especially in the field of plasticity, these models are often developed on the basis of existing gated recurrent units (GRUs) or long-short term memory (LSTM) cells, which have a proven capability to learn path-dependent phenomena. Major drawbacks of these architectures, however, include the relatively long training times and a time-resolution dependent prediction that extrapolates poorly. Further, most existing work for macro- or mesoscopic surrogate models in plasticity is content to predict relatively simple material behaviors. Hence, in this work, the focus lies on circumventing these drawbacks. We formulate a time-resolution independent neural architecture on the basis of the recurrent neural operator and apply it to learn the temperature-dependent plastic response of magnesium, known for its pronounced plastic anisotropy and thermal dependence. The recurrent neural operator predicts the material behavior with great accuracy and is shown to generalize well for various loading cases and temperatures, as well as time-resolutions. Furthermore, it clearly outperforms existing GRU and LSTM models in terms of training and predictive performance. In the second part of this contribution, we perform multiscale simulations with the RNO and achieve at least three orders of magnitude speed-up compared with existing models.  
Neural network surrogate models for constitutive laws in computational mechanics have been in use for some time. In plasticity, these models often rely on gated recurrent units (GRUs) or long short-term memory (LSTM) cells, which excel at capturing path-dependent phenomena. However, they suffer from long training times and time-resolution-dependent predictions that extrapolate poorly. Moreover, most existing surrogates for macro- or mesoscopic plasticity handle only relatively simple material behavior. To overcome these limitations, we introduce the Temperature-Aware Recurrent Neural Operator (TRNO), a time-resolution-independent neural architecture. We apply the TRNO to model the temperature-dependent plastic response of polycrystalline magnesium, which shows strong plastic anisotropy and thermal sensitivity. The TRNO achieves high predictive accuracy and generalizes effectively across diverse loading cases, temperatures, and time resolutions. It also outperforms conventional GRU and LSTM models in training efficiency and predictive performance. Finally, we demonstrate multiscale simulations with the TRNO, yielding a speedup of at least three orders of magnitude over traditional constitutive models.
\end{abstract}{}


\end{frontmatter}
% \tableofcontents
%\linenumbers[5]
% \newpage
% \tableofcontents

\section{Introduction}
\label{sec:Intro}
Plastic behavior is commonly observed in engineering materials and plays a critical role in designing structures that are more durable, stronger, and tougher. This is particularly true for metals, whose plastic deformation behavior is ubiquitous during manufacturing and deformation processes. However, plasticity is a complex phenomenon that is path- and history-dependent, making it challenging to understand and predict. The deformation or loading history can activate a variety of material-specific microstructural mechanisms, which complicates the modeling of metal plasticity.

Crystal plasticity (CP) models have been extensively used to study and model plasticity in metals \citep{Kalidindi1998, ortiz1999variational, staroselsky2003constitutive}, especially in combination with the finite element method (FEM), Fast Fourier Transform (FFT)-based solution schemes, or phase-field and mean-field techniques \citep{Liu2019, Zhang2012, kweon2021study, Ardeljan2016, Tome2000, Beyerlein2004, Tutcuoglu2019a, hollenweger2022efficient}. The major drawback of CP models when used in full-scale numerical simulations is their high computational cost, which becomes even more significant in a multiscale setting. To address this issue, empirical constitutive models or reduced-order models have been developed \citep{Chang2017, Tome2000,Becker2016}, which reduce the full account of all slip (and possibly twin) systems to simpler approximate models with varying degrees of accuracy. As a result, one faces a trade-off between accuracy and computational efficiency when it comes to selecting a plasticity model for simulations (see, e.g., the discussion in~\citet{Chang2017}).
As an alternative, data-driven methods have recently gained attention for overcoming this challenge rooted in the need to repeatedly solve governing equations and boundary value problems in simulation environments and deal with stability issues \citep{zhang2020using,gorji2020potential}. As this is the key focus of our work, we provide a brief overview of this topic in the following. We focus on the micro- and mesoscale representation of polycrystalline metals, where CP models find their application, as they provide a good compromise between accuracy and speed for modeling the microscale response to given loads and deformation. However, our approach is not necessarily limited to that scale, so the discussion can be extended to other relevant (larger and smaller) scales.

\subsection{State of the art}
There are two main approaches to utilizing data-driven techniques for plasticity problems, which exploit machine learning (ML): directly representing the material's constitutive behavior via a neural network (the \textit{direct} approach), and accelerating certain aspects of the computation while keeping the physically inspired structure of the model, e.g. computing crystal plasticity updates but replacing the physical update equations with a neural network (the \textit{indirect} approach).  

Indirect approaches typically use artificial neural networks to learn the relation between certain physical parameters. For instance, \citet{ibragimova2021new} used fully connected neural networks (FCNNs)  in a material model for face-centered cubic (fcc) aluminum (Al) to accelerate the computation of costly constitutive updates, such as the evolution of slip and twin activities or the hardening laws in the CP model. Their setup used an artificial neural network (ANN) to map crystal plasticity model's information at each time step, such as texture, hardening factors, strain rate and critical resolved stresses of the slip systems to the stresses, updated texture and internal variables at the next time step. This model effectively replaces the evaluations of the CP update laws with an ANN. Since conventional CP models often exhibit stability-related constraints in performing these updates, using a neural network to learn the evolution laws can be valuable to circumvent these limitations and to improve efficiency. At the same time, neural networks are enormously fast to evaluate, as they involve highly parallelizable computations, in contrast to mostly serial computations in physical modeling, rendering the evaluations of the models much faster. 

As a drawback, the training of ML models requires extensive datasets. For example, training the model of \citet{ibragimova2021new} required almost 1.5 million samples --- a vast amount of data that must be generated efficiently. Even though the training is done offline, the costs to generate and possibly improve such datasets can be prohibitive, especially as the learned model is specific to the trained scenario. Adjusting the model to new materials or scenarios requires retraining with potentially huge amounts of data. 

Another successful example of the indirect approach was provided by \citet{sun2021cross}, who used FCNNs to perform multi-scale forming simulations. In their approach, cellular automata crystal plasticity finite element models (CACPFEM) are run at RVE level to generate large numbers of data. These data are used for training and as input to ANNs aimed to learn the varying constitutive model parameters during the forming process. These parameters are then used in J2- plasticity models to predict the constitutive response.

In contrast, the direct approach to learning the constitutive response of a material completely bypasses the need for an explicit physical model by directly learning a representation of the constitutive behavior (e.g., from strain to stress space) via a neural network. The result is, mathematically speaking, a map, that maps measures from one space (the input space) into another space (the results space). For instance, \citet{al2006prediction} used a 3-layer FCNN to predict creep stresses in composites by accounting for temperature $T$, strain components $\epsilon_{ij}$, and time $t$ as input. Their model maps the input directly onto the creep stress components, $\sigma_{ij}^\text{dev}$.
\citet{zhang2020using} later represented von Mises plasticity by a similar neural network by learning the hardening moduli.

Despite the above and related successes, modeling plasticity with FCNNs has severe limitations. Plasticity is a process that depends on previous material history, so that neural networks lacking memory in their architecture require inherent knowledge about the nature of the plastic process or input of the plastic flow and hardening. This limits the capability to generalize these network architectures to other problems. Another challenge arises from the resolution (discretization in time and/or space) and the rate dependence \citet{zhang2020using}, which are hard to generalize so that learned models are typically limited to the scenario chosen for generating the training data.

To directly learn the constitutive behavior of plastic materials over general loading paths, recurrent neural networks (RNNs) have been utilized. They learn to predict a series of states (e.g., stresses or plastic strains) of the material in a recurrent manner, so the new state is predicted from the current and/or previous states. The associated architectures are inspired by other fields, notably natural language processing, which deal with serial inputs and outputs. 

Long short-term memory (LSTM) cells, introduced by \citet{hochreiter1997long}, or gated recurrent units (GRU), introduced by \citet{chung2014empirical}, are commonly employed to model plasticity \citep{mozaffar2019deep, yu2022elastoplastic,abueidda2021deep,gorji2020potential,ghavamian2019accelerating}. For example, \citet{yu2022elastoplastic} showed the capability of LSTM and GRU models to predict plastic behavior for J2-plasticity without the need for the plastic strain as input. Instead, their architecture preserved the required information in its "hidden" state. In another study, \citet{mozaffar2019deep} used GRUs to model the three-dimensional (3D) path-dependent stress in elastoplastic materials, while \citet{gorji2020potential} showed that a GRU-based model can capture the constitutive response of a homogeneous anisotropic hardening 2D material. Further, \citet{abueidda2021deep} captured complex, thermoplastic material behavior. They also noted, however, that their architectures were slow in training and rather inefficient for the task at hand. Therefore, they introduced a temporal convolutional network architecture to increase the efficiency and the prediction accuracy. Finally, \citet{ghavamian2019accelerating} demonstrated that RNNs with LSTM cells can learn the viscoplastic behavior of a Perzyna viscoplasticity model and thus accelerate FE$^2$ computations by bypassing the need for evaluating the plastic constitutive model. In a bid to increase the generality of their surrogate model to multiple strain rates, the authors interpolated between results that were trained at different strain rates. They re-trained the NN on varying strain rates and showed decent agreement with the ground truth. Nevertheless, the temporal and spatial resolution dependence still poses a challenge for many applications. 

In general, the above architectures are not a natural choice to capture physical processes in plasticity, as they were developed for natural language processing tasks. These problems typically contain serial input, such as text or a voice recording, and such series are typically non-Markovian. (\textit{Markovian} refers to a process that fulfills the Markov property, i.e., that the probability of future states of the process depends only upon the present state and not on any past states. A Markovian process hence describes only such processes in which the current state perfectly describes the problem without the need for further information to predict the future. In this spirit, conventional plasticity models often assume that all information describing the constitutive behavior is encoded in a set of internal variables \emph{at a given time}. The entire history of the deformation up to that time is hence stored in those internal (or state) variables, so there is no need for information reaching into the past. This is in stark contrast to applications in natural language processing, where the meaning of a sentence depends on the location of individual words that additionally hold different weights to the overall meaning. In architectures such as LSTMs and GRUs, this led to the development of gate functions, which are trained to store long-past information. In applications for plasticity, however, those additional gates and the trainable weights associated with the capability to store long-past information lead to inefficiencies and long training times \citep{abueidda2021deep, yu2022elastoplastic, bonatti2021one}.

In addition, while it is evident that LSTMs and GRUs are extremely versatile, they struggle with capturing plastic behavior \emph{efficiently} \citep{bonatti2021one}. They require a vast number of parameters, leading to overly complex models that further lack interpretability \citep{bonatti2021one, bonatti2022cp}. Further, the gate functions in GRUs and LSTMs lead to an additional issue, since the gates were developed to deal with sequential input of the same frequency -- word for word. Hence, these architectures only perform well in applications, where the input (e.g., the strain) is uniformly spaced (e.g., constant time or load steps). This is a severe limitation for surrogate modeling, where the solver may need to adaptively update the step size. 

Therefore, customized and specialized models for plasticity have been introduced. A prominent approach was shown in \citet{bonatti2021one}, which is known as the minimal state-cell (MSC) architecture. The MSC architecture directly maps the (plastic) deformation to the (deviatoric) stress and learns the evolution of state variables "on the fly" via GRU-inspired update gates. Furthermore, the MSC approach enables the identification of a set of minimal state variables, from where it draws its name. This feature is a significant step up from the non-interpretable states in LSTMs and GRUs. It was exemplified by \citet{bonatti2021one} in a recent study, where a large amount of data obtained from single-element simulations was utilized to predict the deviatoric stress response of various materials, encompassing different hardening types and capturing the Mullins and Bauschinger effects of the tested material models. The MSC identified the number of state variables required for each material, corresponding to the state variables of the phenomenological models used in training plus the five plastic strain components. The approach was later re-used in \citet{bonatti2022cp} to capture the homogenized response of an fcc material, albeit with inconclusive results pertaining to the number of required internal state variables. In its original form, the MSC approach still depends on the training resolution. This issue was addressed in \citet{bonatti2022importance} in the context of self-consistency in surrogate models. The goal of their work was not to provide temporal resolution independence but to at least guarantee consistency under refinement of the path-sampling frequency along the time series. Their approach to self-consistency (i.e., validity across different input frequencies) relied on constraining the gated transition functions. However, there is, unfortunately, no guarantee that the selected transition function is correct or works generally for a wider range of problems \citep{bonatti2022importance}.

\subsubsection{Problem statement}

Despite the potential shown by the aforementioned data-driven methods in the surrogate modeling of plasticity, their use comes with significant drawbacks and challenges. The primary obstacle is the need for a considerable amount of data to train models accurately, which can be both costly and difficult to obtain -- and the required amount of data increases with the complexity of the material behavior. Moreover, data-driven approaches may struggle to generalize beyond the conditions of the training data, emphasizing the importance of selecting representative datasets carefully. Overfitting the training data can lead to poor model performance when applied to new data, and the accuracy of data-driven models depends on the quality and representation of the data used during training. In addition, traditional architectures like LSTMs and GRUs lack interpretability and are limited in capturing general plasticity. Gate functions are also resolution-dependent. While the MSC architecture offers some relief from these issues, resolution independence is not guaranteed and has only been achieved for plasticity models that were significantly simpler than the anisotropic crystal plasticity models , e.g., to describe hcp metals like magnesium (Mg). 

More generally, the vast majority of ML-surrogate models so far have focused on either capturing the homogenized, often isotropic, response of a material or on learning the behavior of phenomenological constitutive models with limited physical accuracy. Such models do not require a large state space and are often described via an equivalent plastic strain measure $\epsilon_\text{p}$ \citep{bonatti2021one, mozaffar2019deep} or simple hardening laws \citep{ghavamian2019accelerating}. More recently, CP-RNN surrogates were introduced for single- and polycrystals of fcc materials \citet{bonatti2022cp}. This more complex case showed limitations in the MSC architecture's capability to identify an exact number of state variables for more involved formulations of physical models. This is to be expected, as the models initially learned by the MSC cell had a direct connection with the evolution of the plastic stresses and the plastic strain components \citep{bonatti2021one}.

\begin{figure}
    \centering
    \includegraphics[width=0.9\textwidth]{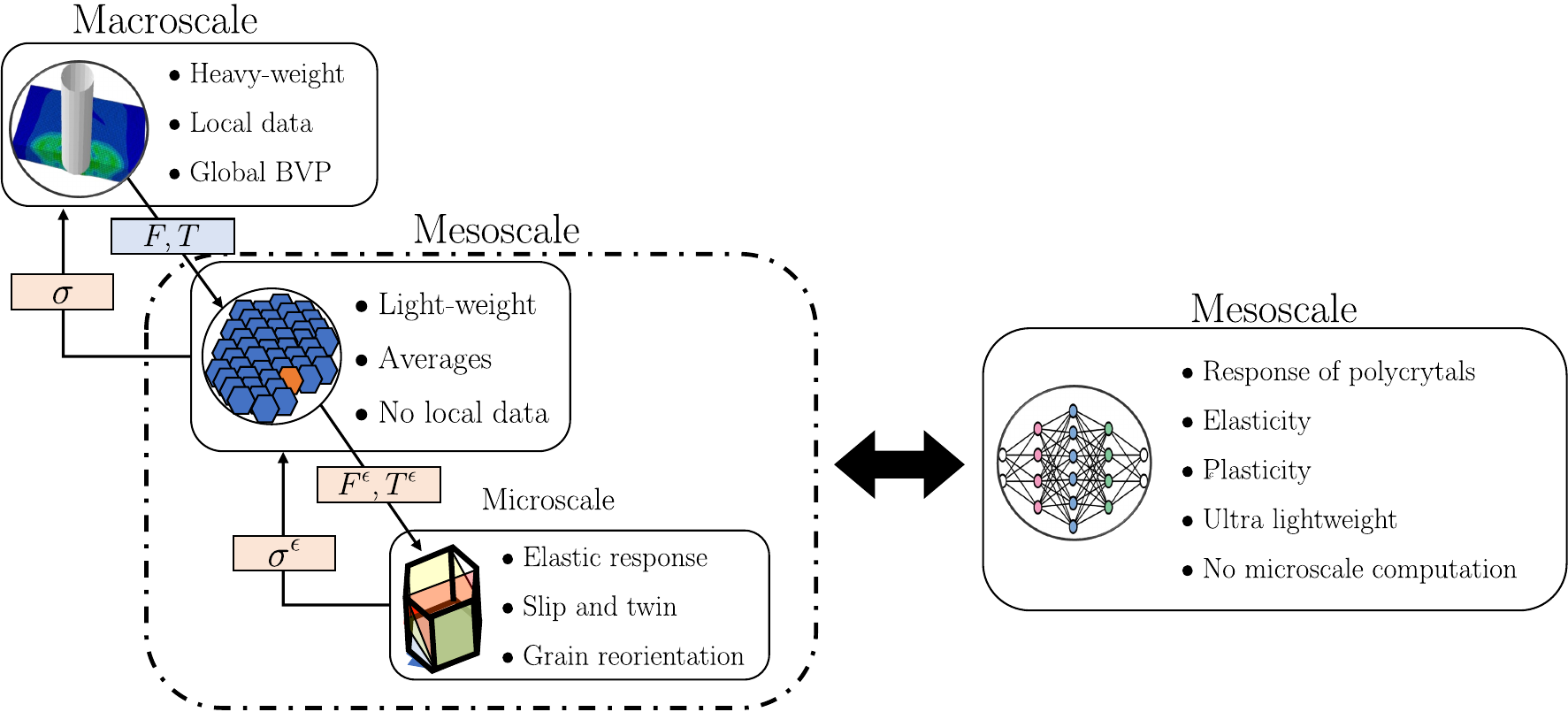}
    \caption{Schematic of the \emph{direct approach}, utilizing a mesoscopic surrogate model to represent the effective constitutive response of polycrystals. }
    \label{fig:Surrogate}
\end{figure}

\subsection{Proposed solution}

In this study, we address the above challenges and propose a new ML-based approach with a two-fold objective. First, we aim to seamlessly integrate self-consistency into the model formulation without the need for cumbersome modulation functions. To this end, we build upon the approach proposed in \citep{bhattacharya2022learning}, which is based on a Markovian descriptor and consists of identifying a set of internal variables ${\xi}$ as well as two functions, $\mathcal{F}$, and $\mathcal{G}$, capable of representing the constitutive model. The first, hidden (or inaccessible) function $\mathcal{G}$ is designed to learn the \emph{rate of change} of ${\xi}$. The new state of the system is then found via discrete time integration, e.g., via an Euler-forward finite-difference scheme. The resulting map leads to a description in continuous time, meaning that the learned functional map has general validity and shows resolution independence, naturally leading to self-consistency based on the new state. In conjunction, $\mathcal{F}$ is trained to predict the final value of interest, e.g., the stress response of the material. This approach, in essence, mimics crystal plasticity models based on the internal-variable theory \citep{rice1971inelastic}. A related framework was used in a previous study \citep{LIU2022104668} based on a Markovian RNN, which showed self-consistency of the recurrent neural operator (RNO) under the tested conditions. 

Second, we aim to learn \textit{more complex mechanical behavior} at the micro- or mesoscale than prior surrogate models. As an example of complicated plasticity models, we consider Mg, which has the potential to become an important player in aerospace and automotive applications \citep{Nie2020}. Notably, it shows pronounced plastic anisotropy (even at the polycrystalline level) and is governed by the competition between slip and twinning \citep{Nie2012, Nie2020}. Modeling Mg-based materials presents a challenging task in computational plasticity \citep{hollenweger2022efficient}. In addition, Mg exhibits strong temperature dependence, which substantially alters its mechanical response and degree of isotropy. The thermo-mechanical interplay of Mg's deformation modes is not fully understood to date \citep{Beyerlein2004, jain2007modeling, Roy2017} but essential for the optimization and design of Mg-based materials \citep{eswarappa2021strengthening}.

% Therefore, in this study we explore the potential of RNO architectures to act as surrogate models for strongly anisotropic materials across a range of temperatures, taking the example of pure Mg. We develop and train an RNO architecture and compare its performance in terms of quality and training with established GRU and LSTM architectures for plasticity. We further demonstrate, the performance of the trained RNO within commercial FEM routines to perform temperature-dependent multiscale simulations of dynamic impact. The novelty of this study is the direct learning of temperature-aware and time-step-independent constitutive responses from stress-strain data in a material with anisotropic temperature-dependent plasticity. We thus investigate the potential benefits of designing physically informed and customized architectures to accurately model complex material behavior. 
In this study, we develop temperature-dependent recurrent neural operators (TRNOs) capable of approximating the constitutive behavior of strongly anisotropic materials, such as pure magnesium, across a range of temperatures. We design and train a TRNO architecture, and compare its performance in terms of accuracy and training efficiency with established GRU and LSTM architectures for modeling plasticity. Additionally, we demonstrate the integration of the trained TRNO into commercial finite element method (FEM) routines to conduct temperature-dependent multiscale simulations of dynamic impact. The novelty of this work lies, among others, in the direct learning of temperature-aware, time-step-independent constitutive responses from stress-strain data for materials exhibiting anisotropic, temperature-dependent plasticity. This study explores the advantages of designing physically informed, tailored architectures to accurately model complex material behavior. Though focused on magnesium in this study, the presented approach is sufficiently general to extend to other types of materials and constitutive models.

\section{Background} 
\label{sec:Background_CP}
Before delving into the data-driven model we provide a brief overview of the material and its unique characteristics, as well as a short review of the physical plasticity model used for the generation of the training data. More details on the material behavior can be found in \citet{Nie2020} and the model is described in depth in \citet{hollenweger2022efficient}.

\subsection{Plasticity of magnesium}
\label{sec:material behavior}

The hexagonal close-packed (hcp) crystal structure of Mg leads to plastic deformation being accommodated by the combined activation of slip and twin mechanisms. These include the basal, prismatic, and pyramidal slip systems, as well as the tensile and compressive twin systems (TT and CT, respectively). Notably, all these systems exhibit significantly different critical resolved shear stresses (CRSSs) implying distinct activation barriers. This, in turn, leads to pronounced anisotropy and a complex plastic deformation behavior, as shown numerically \citep{Zhang2012, agnew2005plastic} and experimentally \citep{Kelley1967, Akhtar1969, yoo1967slip}.

The basal slip systems are widely accepted as the easiest to activate, followed by the prismatic $\langle a \rangle$ and pyramidal $\langle c+a \rangle$ systems, which have orders-of-magnitude higher CRSS values than the basal system \citep{yoo1967slip, yoo2002nonbasal, yoshinaga1964nonbasal}. However, the basal systems alone cannot provide sufficient deformation modes to facilitate general plastic flow \citep{sandlobes2011role, agnew2001, staroselsky2003constitutive}, so that the activation of non-basal slip systems is required to obtain five independent deformation modes, necessary for general plastic deformation \citep{Kelley1967, lilleodden2010microcompression, taylor1938plastic}. 
Non-basal slip, however, has CRSS values one (or several) orders of magnitude higher than the basal system \citep{Chapuis2011, Akhtar1969, yoshinaga1964nonbasal,yoo1967slip,yoo2002nonbasal,lilleodden2010microcompression}. This leads to a lack of \emph{readily activatable} deformation modes in Mg. As a consequence, the material shows twinning in addition to slip to accommodate plastic deformation. Twinning involves the mirroring of a region inside a crystal along a given mirror plane, leading to a reorientation of the underlying lattice \citep{Kelley1967}. This reorientation enables the accommodation of plastic deformation. %typically along the $ \langle c+a \rangle$ direction of the crystal. 

Upon twinning, further deformation is accommodated by slip on the reoriented slip systems. As long as the twins propagate, the constitutive response is controlled by the soft response associated with the twin's lateral expansion \citep{Kelley1967,Wonsiewicz1967,jain2007modeling}. However, once the twins saturate in a crystal, the lattice is entirely reoriented and the rotated lattice's slip systems govern the material's subsequent response. This results in significant changes in the constitutive response, experimentally shown in single crystals for instance in \citep{Kelley1967}. 

\begin{figure}[h]
    \centering
    \includegraphics[width=0.7\textwidth]{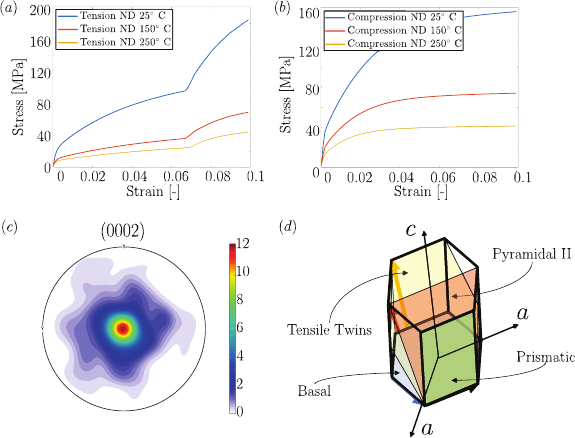}
    \caption{Simulated stress-strain response of a strongly textured polycrystal under (a) tension and (b) compression along the normal direction (ND). The texture of the modeled sample is shown in (c), and the main deformation modes of Mg in (d).}
    \label{fig:Polycrystal_Stress}
\end{figure}

This behavior is showcased in Fig. \ref{fig:Polycrystal_Stress}(a) for a strongly textured Mg polycrystal obtained from a model that was calibrated on Mg single crystal data \citep{hollenweger2022efficient}. While this behavior is more pronounced in single crystals, it is also a common phenomenon in polycrystals, where the severity of the behavior increases with the strength of the texture \citep{hollenweger2022efficient, jain2007modeling, Ardeljan2016, Kelley1967}. In this rolled sample, the majority of the grains' $c$-axes are aligned with the normal direction (ND) of the rolling process (see Fig.~\ref{fig:Polycrystal_Stress}(c)). When loaded in tension, an initial elastic response is followed by pronounced twinning, whose low CRSS and hardening leads to low stress levels. At approximately 6\% strain, twinning is complete and the crystal has reoriented to a configuration in which basal slip is unfavorable, so that the stiffer prismatic and $\langle c+a \rangle$ pyramidal slip systems become the only viable options for plastic flow. This results in the shown stiffening after the saturation of the twins.

By contrast, the response in compression does not show such behavior, which is due to the highly directional nature of twinning \citep{Yoo1981}. 
% We distinguish between compressive and tensile twins (TTs) who both occur under various circumstances. In the present work, 
The tensile twin (TT) systems lead to an extension of the $c-$axis of the crystals and hence are active only under $c-$axis extension. As a consequence, this deformation mode is inactive when the material is compressed along the $c-$axis, as evident in Fig.~\ref{fig:Polycrystal_Stress} (b), which does not show any twin-induced changes in the stress-strain curve. Instead, the material accommodates deformation primarily through the stiff $\langle c+a \rangle$ pyramidal slip, which exhibits strong strain hardening. (In reality, such polycrystals show signs of failure at an applied strain of approximately 5-8\% \citep{Kelley1967, Wonsiewicz1967}.) 

As a consequence of the complex microstructural deformation modes in Mg, its constitutive behavior exhibits a strong directional tensile and compressive anisotropy. 

Further, the material behavior exhibits significant temperature dependence\citep{Chapuis2011, Akhtar1969, Kelley1967}, owing to changes in CRSS and hardening behavior of the twin and slip systems with temperature. Fig.~\ref{fig:Polycrystal_Stress}(a,b) shows stress-strain curves for a range of temperatures. It was noted by \citet{Chapuis2011, jain2007modeling} that TTs persist to a large extent even at elevated temperatures, whereas the reoriented slip systems exhibit a considerably weaker response due to thermal softening. 

In \citet{hollenweger2022efficient}, we developed a crystal plasticity (CP) model to capture the material's complex constitutive behavior across a wide temperature range. Here, we leverage that model to generate training data for the RNO model. Due to the complexity of the material behavior, the CP model must account for the contributions of the varying slip systems as well as the twin reorientation, the temperature dependence of the slip systems, and the orientation dependence of the plastic response. Since the CP model was described in detail in \citet{hollenweger2022efficient}, we here offer a brief description of the model to the extent necessary for subsequent discussions. 

\subsection{Single-crystal plasticity model: kinematics} \label{sec:model_singleCP}

We use the conventional multiplicative split of the deformation gradient $\bm F$, according to \citet{Kalidindi1998, asaro1977strain}, i.e., 
\begin{equation}
   \bfF = \bfF_\text{e}\bfF_{\text{in}},
\end{equation}
with the elastic contribution $\bfF_{\text{e}}$ and the inelastic contribution $\bfF_{\text{in}}$. The latter arises from the combined slip and twin activity in the material. 
The total velocity gradient follows as
\begin{equation}
\label{eq:VelocityGradient}
\bfl = \dot{\bfF} \bfF^{-1} 
= {\bfl}_\text{e}+ {\bfF}_\text{e}{\Tilde{\bfl}}_\text{in}{\bfF}^{-1}_\text{e}, \qquad
\Tilde{{\bfl}} = {\Tilde{\bfl}}_{\text{p}}+{\Tilde{\bfl}}_{\text{tw}}
\end{equation}
where ${\bfl}_\text{e}$ and ${\Tilde{\bfl}}_\text{in}$ represent the elastic and inelastic contributions, respectively. The inelastic contribution is further divided into slip and twinning, differentiating between the two plastic deformation mechanisms. In the present model we consider $n_\text{s} = 12$ slip systems \citep{hollenweger2022efficient}: the three basal, three prismatic $\langle a \rangle$, and six pyramidal $\langle c+a \rangle$ systems. In addition, we consider the presence of $n_\text{tw} = 6$ tensile twin systems. This representation of plasticity corresponds to the reduced model version of \citet{hollenweger2022efficient}.

The effect of twinning is assumed of the form
\begin{equation}
{\Tilde{\bfl}_{\text{tw}}} = \sum_{\beta = 1 }^{n_\text{tw}} \dot{\lambda}_{\beta} \gamma^{\text{tw}}_{\beta} {\bfa_{\beta} \otimes \bfn_{\beta}},
\end{equation}
which includes the $\{ 10\bar12\}$ tensile twin systems.
$\gamma^{\text{tw}}_{\beta}$ is the twinning shear associated with twin system $\beta$, and $\bm a_\beta$ and $\bm n_\beta$ are the twin direction and normal of the twin plane, respectively. 
% In Mg, for the common $\{ 10\bar12\}$ tensile twin systems, $\gamma^\text{tw}_\beta $ is as a material constant known from crystallography \citep{Zhang2012, Yoo1981}. 

The slip contribution for the velocity gradient is of a different form, owing to the different nature of the evolution of the slip systems. Slip occurs in the parent material as well as in the twinned regions of material. Hence, the evolution of plastic deformation from slip has to account for both components. It reads 
\begin{equation}
\label{eq:l_plastic}
    {\Tilde{\bfl}}_{\text{p}} = \sum_{\alpha = 1}^{N_\text{s}} \dot{\gamma}_{\alpha} \underbrace{\left[
    \left(1-\sum_{\beta=1}^{N_\text{tw}} \lambda_{\beta} \right) {\bfs}_{\alpha}\otimes{\bfm}_{\alpha} + \sum_{\beta=1}^{N_\text{tw}} \lambda_{\beta} \bfs'_{\alpha}\otimes\bfm'_{\alpha}
    \right]}_{={\bfp_\alpha}},
\end{equation}
where the first contribution represents the slip  in the parent material. The fraction of the parent material is given as one minus the sum of all twin fractions in the grain and represents a volume fraction. Here, $\dot{\gamma}_{\alpha}$ denotes the slip rate of slip system $\alpha \in [1,n_\text{s}]$, which is described by slip direction ${\bfs}_{\alpha}$ and slip plane normal ${\bfm}_{\alpha}$ in the undeformed material.  
The second component of the sum in \eqref{eq:l_plastic} indicates the deformation in the twinned region. We represent the reoriented slip systems in the twined regions following the approach of \citet{christian1995deformation}:
\begin{equation}
{\bfs}'_{\alpha} = {\bfQ}_{\beta} {\bfs_\alpha}, \ {\bfm}'_{\alpha} = {\bfQ}_{\beta} {\bfm_\alpha}
\st{with}
{\bfQ_{\beta}} = {I} - 2 {\bfn_{\beta} \otimes \bfn_{\beta}},
\end{equation}
where $\bfQ_\beta$ is a rotation matrix, representing the local reorientation of the lattice due to twin system $\beta$.

\subsection{Single-crystal plasticity model: constitutive relations}

We define the overall Helmholtz free energy density as
\begin{equation}
\label{eq:Overall Helmholtz}
W({\bfF,\bfF_{\text{in}}, \bfepsilon, \bflambda},T) = W_{\text{e}}({\bfF}_{\text{e}},T) + W_{\text{p}}({\bfepsilon,\bflambda},T) + W_{\text{tw}}({\bflambda},T),
\end{equation}
which contains an elastic strain energy density $W_{\text{e}}$ along with energy contributions $W_{\text{p}}$ and $W_{\text{tw}}$ accounting for hardening due to slip and twinning, respectively. $T$ is the temperature. Associated with each twin or slip system, we introduce a set of internal variables representing their accumulated activity $\bm\gamma = (\gamma_1 \dots, \gamma_{n_\text{s}})$ and twin volume fractions $\bm\lambda = (\lambda_1 \dots, \lambda_{n_\text{tw}}) $, respectively. 

The plastic hardening of the twins is straightforward and depends on the volume fraction of the twin systems. The twin hardening description is based on the energy density \citep{hollenweger2022efficient}
\begin{equation}
\label{eq:TwinHarden}
    W_{\text{tw}} = \underbrace{\sum_{\beta = 1}^{n_t} \frac{k_{\beta}(T)}{2}{\lambda}^{2}_{\beta}}_{\text{self-hardening}}  + \underbrace{\frac{1}{2} {\bflambda} \cdot \mathcal{K}(T){\bflambda} }_{\text{cross-hardening}}.
\end{equation}
Here $k_\beta (T)$ is a temperature-dependent self-hardening factor, indicating how much the twin system hardens \emph{itself}, whereas $\mathcal{K}$ denotes the cross-hardening between different twin systems. $\mathcal{K}$ is a positive semi-definite hardening matrix with zero diagonal and non-zero twin cross-hardening terms $k_{\beta\beta'}$ on its off-diagonals. Slip-twin hardening is not considered. Also, note that only the tensile twin variant with the highest critical resolved shear stress evolves at one time, hence this factor plays a subordinated role in the absence of the compressive twin systems. In addition, double-twinning is not included at this stage of the model\citep{hollenweger2022efficient}. Hence, once one of the tensile variants has reached a saturation value $\lambda_\text{crit} = 85\%$, the crystal is considered as "fully twinned" and the reorientation is assumed to be complete. 

For the slip systems, the hardening theoretically depends on the accumulated dislocation density. Since the CP model does not track these, instead we introduce the total plastic activity on a slip system ${\bfepsilon}=(\epsilon_1,\dots,\epsilon_{N_\text{s}})\T$, evolving with the plastic slip rates according to $\dot\epsilon_\alpha=|\dot\gamma_\alpha|$. This ensures that any type of plastic deformation on the slip system leads to strain hardening. The slip-hardening energy therefore reads as:
\begin{equation}\label{eq:Wp}
W_{\text{p}} =
\begin{cases}
        & \underbrace{\sum_{\alpha = 1}^{n_s}\sigma^{\infty}_\alpha(T) \left[ \epsilon_{\alpha} + \frac{\sigma^{\infty}_\alpha(T)}{{h^{0}_\alpha(T)}}\exp\left( - \frac{h^0_\alpha(T) \epsilon_\alpha}{\sigma^{\infty}_\alpha(T)} \right)\right]}_{\text{self-hardening}}
         \quad +  \underbrace{\frac{1}{2} {\bfepsilon} \cdot \mathcal{H}(T){\bfepsilon} }_{\text{cross-hardening}},
    \quad \text{ if $\lambda < \lambda_{\text{crit}}$}, \\
       & \underbrace{\sum_{\alpha = 1}^{n_s}\sigma^{\infty}_\alpha(T) \left[ \epsilon_{\alpha} + \frac{\sigma^{\infty}_\alpha(T)}{{h^{0}_\alpha(T)}}\exp\left( - \frac{h^0_\alpha(T) \epsilon_\alpha}{\sigma^{\infty}_\alpha(T)} \right)\right]}_{\text{self-hardening}} 
        \quad +  \underbrace{\frac{1}{2} {\bfepsilon} \cdot \mathcal{H}(T){\bfepsilon} }_{\text{cross-hardening}} +
        \underbrace{\sum_{\alpha=1}^{N_\text{s}} c_
        \alpha(T)\epsilon_\alpha}_{\text{twin-slip interaction}},
    \text{ if $\lambda \geq \lambda_{\text{crit}}$},
\end{cases}
\end{equation}
where $\lambda$ refers to the volume fraction of the primary twin system. $\mathcal{H}(T)$ is a positive semi-definite hardening matrix and represents slip-slip cross-hardening through the off-diagonal terms $h_{\alpha\alpha'}$ (again having zero diagonal).  $\sigma_\alpha^\infty$, $h_\alpha^0$, and $c_\alpha$ are temperature-dependent model parameters that are summarized in \citet{hollenweger2022efficient}. All material parameters used in subsequent simulations are summarized in Tables~\ref{tab:RT calibrated values} and \ref{tab:RT calibrated values2} for the slip and twin systems, respectively.

\begin{table}
    \centering\small
        \resizebox{\columnwidth}{!}{
    \begin{tabular}[width=\textwidth]{lllllll}
       \textbf{Parameter} & \textbf{Symbol} &  \textbf{Unit}  & \textbf{Basal} & \textbf{Prismatic} &  \textbf{Pyramidal II} &
       \textbf{Reference/Calibration} \\
        \hline
        CRSS & $\tau_0$ & MPa & 2.0 & 20 & 40 ($100^{\dagger}$) & calibration$^{*}$\\
        Self hardening factor & $h_0 $  & MPa &  750 & 6000 & 13000  & calibration\\
        Cross hardening factor & $h_{ij}$ & MPa  & 10 & 10 & 12 & calibration \\
        Saturation stress & $\sigma^{\infty}$ &MPa & 0.8 & 72 & 115 ($60^{\dagger}$) & calibration\\
       Twin-slip cross-hardening factor & $c_{\alpha}$ &MPa & 20 & 10 & 5 & calibration\\
       Reference slip rate & $\dot{\gamma_0}$ & $s^{-1}$ & $10^{-3}$ &  $10^{-3}$ &  $10^{-3}$ & \citet{Zhang2012}\\
       Slip exponent & $m_{p}$ & -- & 0.5  & 0.5 &  0.5 & \citet{Chang2015}\\
       CRSS temperature factor & $\omega$ & -- & -- & 3.0 & 2.95 ($5.5^{\dagger}$) & calibration$^{*}$
       \\
       Self-hardening temperature factor & $\eta$ & -- & -- &  5.2 & 4.5  ($2.2^{\dagger }$) & optimization
       \\
       Cross-hardening temperature factor & $\nu$ & -- & -- & 1.3 & 2.5 & optimization
       \\
       Saturation stress temperature factor & $\xi$ & -- & -- & 2.5 & 4.05 ($2.5^{\dagger }$) & optimization
    \end{tabular}
    }
    \caption{Model parameters for the slip systems at room temperature, calibrated based on the experimental data of \citet{Kelley1967}. Values marked by an asterisk ($^*$) were calibrated based on data from \citet{Chapuis2011,Akhtar1969,lilleodden2010microcompression,ando2007temperature,Kelley1967}. The temperature dependence of the CRSS values was calibrated such as to agree well with experimentally reported data \citep{hollenweger2022efficient}.}
    \label{tab:RT calibrated values}
\end{table}

\begin{table}[h!]
    \centering
    \resizebox{\columnwidth}{!}{
    \begin{tabular}[width=\textwidth]{llllll}
       \textbf{Parameter} & \textbf{Symbol} &  \textbf{Unit}  & \textbf{TT} & \textbf{CT} & \textbf{Reference/Calibration} \\
        \hline
        CRSS & $\tau_0$ & MPa & 3.5 & 50  & calibration\\
        Self hardening factor & $k_0 $  & MPa &  25 & 3000 & calibration\\
        Twin shear & $\dot{\gamma_0}$ & -- & 0.129 & 0.138 & \citet{Zhang2012, Chang2015} \\
        Cross hardening factor & $h_{ij}$ & MPa  & 100 & 100 & \citet{Chang2015}\\
        Critical volume fraction & $v_{\text{crit}}$ &-- & 0.9 & 0.9 & \citet{Zhang2012} \\
        Twin rate & $\dot{\gamma_0}$ & -- & 10$^{-3}$ & 10$^{-4}$  & \citet{Zhang2012}\\
        Twin exponent & $m_{\text{tw}}$ & -- & 1 & 1 & \citet{Chang2015} \\
        CRSS temperature factor & $\omega$ & -- & -- & 3.5 & calibration\\
        Self-hardening temperature factor & $\xi$ & -- & -- & 4.0 & calibration 
    \end{tabular}
    }
    \caption{Model parameters for the twin systems, calibrated based on the experimental data of \citet{Kelley1967} and \citet{Wonsiewicz1967}.}
    \label{tab:RT calibrated values2}
\end{table}

Assuming that both slip and twinning follow a Schmid-type law, the resolved shear stress (RSS) of slip system $\alpha$ is expressed as 
\begin{equation}
    \tau_{\alpha}^\text{p} = {\bfSigma} \cdot {\bfp}_{\alpha},
\end{equation}
With the Mandel stress tensor defined as \begin{equation}{\bfSigma} = {\bfF}{\text{e}}\T{\bfP\bfF}{\text{in}}\T, \end{equation} based on the first Piola Kirchhoff stress tensor ${\bfP} = \partial W/\partial {\bfF}$. The twin RSS follows analogously as
\begin{equation} 
    \tau_{\beta}^\text{tw} = {\bfSigma} \cdot \left(\gamma_{\beta}^{\text{tw}}{\bfa}_{\beta} \otimes {\bfn}{\beta}\right)
\end{equation}

The evolution laws, governing the rate of change of the internal model variables (slip activity and twin volume fraction),  are shown in \citet{hollenweger2022efficient}. For slip, the law reads
\begin{equation}
\label{eq: slip Update}
    \dot{\gamma}_\alpha = \dot{\gamma}_{0,\alpha} \left \lvert \frac{|\tau^{\text{p}}_\alpha| - \frac{\partial W_{\text{p}}}{\partial\epsilon_\alpha}}{\tau_{0,\alpha}(T)} \right\lvert  ^{1 / {m_{\text{p}}}} \text{sign}(\tau^{\text{p}}_\alpha), 
\end{equation} 
with reference slip rates $\dot{\gamma}_{0,\alpha}$, a constant. Further, a temperature-dependent CRSS of the slip systems $\tau_{0,\alpha}(T)$ was used, following experimental results from \citet{Chapuis2011,Akhtar1969,Kelley1967,Wonsiewicz1967}. Finally, the hardening law follows an exponential form with hardening exponent $m_\text{p}$. Since slip systems can experience movement along the positive or negative slip direction, the evolution law depends on the sign of the RSS. 

The twin volume fractions follow a similar pattern
\begin{equation}
\label{eq:update}
        \dot{\lambda}_\beta = \dot{\lambda}_{0,\beta} \left \lvert \frac{|\tau^{\text{tw}}_\beta| - \frac{\partial W_{\text{tw}}}{\partial\lambda\beta}}{\tau_{0,\beta}} \right\lvert ^{1 / {m_{\text{tw}}}}
\end{equation}
with a reference slip rate $\dot\lambda_{0,\beta}$, hardening exponent $m_\text{tw}$, and temperature-independent CRSS $\tau_{0,\beta}$. Since twins are directional and can only form along the positive twin direction, no sign of $\tau^\text{tw}_\beta$ is needed. 

The constitutive model is completed by a choice of the elastic energy density, for which we here assume takes the simple Neo-Hookean form \citep{Chang2015,hollenweger2022efficient}
\begin{equation}
    W_{\text{e}} ({\bfF}_{\text{e}},T) = \frac{\mu(T)}{2}(\tr\bar{{\bfC}}_{\text{e}} - 3) + \frac{\kappa(T)}{2}(J - 1)^{2}
\end{equation}
with temperature-dependent shear modulus $\mu(T)$ and bulk modulus $\kappa(T)$. Further the present formulation of the Neo-Hookean is based off the normalized right Cauchy-Green tensor
\begin{equation}
    \bar{{\bfC}}_{\text{e}} = \bar{{\bfF}}^{T}_{\text{e}} \bar{{\bfF}}_{\text{e}} , \quad { \text{where }\bar{{\bfF}}_{\text{e}}} =  \frac{{\bfF}_{\text{e}}}{J^{1/3}}, \quad J = \det{\bfF_\text{e}}=\det{\bfF}.
\end{equation}
Of course, this is a simplified choice that neglects the elastic anisotropy, yet the elastic strains are comparably small in our study, so that the exact form of $W_\text{e}$ plays only a minor role in the following (see also the discussions in \citep{Chang2015,Chang2017}).

\subsection{Temperature dependence}
The temperature dependence of the material is modeled by an Arrhenius-type law for the temperature dependence of the CRSS values \citep{Beyerlein2004,Liu2017,Wang2019}. The temperature dependence of the slip systems is assumed to decay exponentially with temperature, i.e.,
\begin{equation}
    \tau_{0,\alpha} (\bar{T}) = 
    \begin{cases}
        \tau_{0,\alpha}(T_\text{ref}), & \text{if $\alpha$ is a basal system}, \\
        \tau_{0,\alpha}(T_\text{ref}) \exp(- \omega_{\alpha}\bar{T}) & \text{else}
    \end{cases} 
\end{equation}
with distinct parameters $\omega_\alpha>0$ for the prismatic and pyramidal systems, and the dimensionless temperature
\begin{equation}
    \bar{T}(T) = \frac{T - T_{\text{ref}}} {T_{\text{melt}}-T_{\text{ref}}}.
\end{equation}
Here $T_{\text{melt}} = 650^{\circ}C$ is the melting temperature of Mg, and we define the reference temperature as $T_{\text{ref}}=25^{\circ}C$. The basal systems are ssumed as temperature-independent following insights from \citet{Chapuis2011, Wonsiewicz1967, Akhtar1969, Liu2017}. 

The same ansatz is used to describe the temperature sensitivity of the hardening parameters \citep{hollenweger2022efficient}:
\begin{equation}
\begin{split}
h^{0}_\alpha(\bar{T}) = &
\begin{cases}
   h^{0}_\alpha(T_\text{ref}) & \text{if $\alpha$ is a basal system},\\
    h^{0}_\alpha(T_\text{ref})\exp(-\eta_\alpha \bar{T}) & \text{else},
\end{cases} \\
 h_{\alpha\alpha'}(\bar{T}) = &
\begin{cases}
   h_{\alpha\alpha'}(T_\text{ref}) & \text{if $\alpha$ is a basal system},\\
   h_{\alpha\alpha'}(T_\text{ref})\exp(-\chi_\alpha \bar{T}) & \text{else},
\end{cases} \\
\sigma^{\infty}_\alpha(\bar{T}) = &
\begin{cases}
   \sigma^\infty_\alpha(T_\text{ref}) , & \text{if $\alpha$ is a basal system}, \\
    \sigma^\infty_\alpha(T_\text{ref})\exp(-\nu_\alpha \bar{T}) & \text{else}.
\end{cases}
\end{split}
\end{equation}
Analogously, the hardening laws for TTs are assumed to be temperature insensitive:
\begin{equation}
k^{0}_\beta(\bar{T}) =
\begin{cases}
   k_\beta(T_\text{ref}) & \text{if $\beta$ is a TT system}, \\
   k_\beta(T_\text{ref}) \exp(-\xi_\beta \bar{T}) & \text{if $\beta$ is a CT system}.
\end{cases}
\end{equation}
Finally, the elastic constants are chosen as $\mu(T) = 17 \exp(-3\bar{T})$ GPa and $\kappa(T) = 16 \exp(-3\bar{T})$ GPa, by extending the (room-temperature) Neo-Hookean formulation of \cite{Vidyasagar2018}. 

\subsection{Polycrystalline response} 

In this paper, we focus on the polycrystalline behavior of Mg. A polycrystal is a medium made of a collection of grains, where we here assume a homogeneous material in terms of composition and grain size. The texture of the polycrystals under consideration is that of a pronounced rolling texture in pure Mg similar to the one reported in  \citet{Kelley1967, hollenweger2022efficient}. The response of a Mg polycrystal results from the homogenized response of this ensemble of grains. 

Numerically, numerous options for solving this homogenization problem exist, such as full-field methods like the FEM \citep{Zhang2012,Indurkar2020,Liu2017} or spectral methods \citep{Moulinec1998,Vidyasagar2018,Tutcuoglu2019a}. For large-scale data generation, however, these approaches take too long. Mean-field methods, on the other hand, such as the Taylor model (based on \citet{taylor1938plastic}'s assumption of isostrain plasticity) or the viscoplastic self-consistent model \citep{Tome2000} can rapidly compute ensemble averaged responses of the polycrystals. For the present application, namely as a tool to generate large amounts of training data, the Taylor model is the best option, due to its ease of parallelization and simplicity. 

\subsection{Taylor approach to polycrystal modeling}
\label{sec:Taylor_CP}

The polycrystal response is obtained from the single-crystal CP model by invoking the Taylor assumption of equal strains across all grains and averaging according to the size of the grains. This approach yields an upper bound on the stress\citep{hollenweger2022efficient}, meaning, it has tendency to over-approximate the real response. This problem can, however, be alleviated via relaxing the Taylor assumption, as shown in \citet{Chang2015}. 

Mathematically, the polycrystal response results from differentiating an averaged strain energy density, given by the average of the individual grain's energy densities, with respect to the deformation gradient.
This yields the first Piola-Kirchhoff stress tensor of the polycrystal, as an ensemble average,
\begin{equation}
       {\bfP} =  \frac{\partial W_\text{Taylor}}{\partial {\bfF}},
\end{equation}
while the Cauchy stress tensor follows as
\begin{equation}
\label{eq:Nanson_stress}
    {\bfsigma} = J^{-1} {\bfP} {\bfF}\T.
\end{equation}

For averaging, $f_i$ is introduced as the volume fraction of grain $i$ so that the energy density of the grain ensemble reads
\begin{equation}
    W_\text{Taylor} = \sum_{i = 1}^{N} f_i W_i({\bfF},{\bfepsilon_i, \bflambda_i}, T).
\end{equation}
Each grain further possesses an orientation wrt. the global coordinate system described by the (initial) rotation tensor ${R}_i\in\mathrm{SO}(3)$.
This rotation defines the positioning and orientation of the slip and twin systems in grain $i$ via
\begin{equation}
\label{eq:TaylorGrains1}
% \begin{split}
 {\bfs}_{i,\alpha} = {\bfR}_i {\bfs}_{\alpha}, \qquad
 {\bfm}_{i,\alpha} = {\bfR}_i {\bfm}_{\alpha},
% \end{split}
\end{equation}
and 
\begin{equation}
\label{eq:TaylorGrains2}
% \begin{split}
 {\bfn}_{i,\beta} = {\bfR}_i {\bfn}_{\beta}, \qquad
 {\bfa}_{i,\beta} = {\bfR}_i {\bfa}_{\beta}.
% \end{split}
\end{equation}

% The physical internal variables, which include the accumulated shear on the slip systems and the volume fraction of the twin systems, are collectively stored in ${{\eta}} = [{\epsilon,\lambda}]$, for each grain. In the Taylor model, all grains undergo the same deformation gradient ${F}$. The grains do not impact each-other during deformation, as would e.g. be the case in an FE model. Therefore, the individual grain response only depends on ${F}$ and the stored internal variables, differing for each grain due to the individual grain orientations. 

% The Cauchy stress can further be decomposed into a hydrostatic pressure $p_\text{hydro}$ contribution and the deviatoric stress, i.e.,
% \begin{equation}
%     {\bfsigma} = \sigma^\text{h}\mathbb{I} + {\bfsigma}^\text{dev}
%     \st{with}
%     \sigma^\text{h} = \frac{\tr{\bfsigma}}{3}, \quad
%     {\bfsigma^\text{dev}} = {\bfsigma} - \frac{\tr{\bfsigma}}{3} \mathbb{I}.
% \end{equation}
% This decomposition allows us to conveniently learn the purely plastic response of the material without the impact of pressure, which be exploited when training our model. 
It should be noted that the presented formulation and homogenization contain a number of implicit assumptions. The first of these lies in the absence of locality on the microstructure. The Taylor model weighs each grain by its volume (or size). Hence a notion of grain neighbors or position of grains within the material is irrelevant and absent. This means that geometric constraints or the balance of forces between grains are absent, leading to stresses that can vary drastically within the grains due to the isostrain assumption. Further, in the presented model uniformity of temperature is assumed across the grains with no heat transfer. This latter assumption generally holds for small strain rates. 

In our simulations, we focus on the temperature-dependent constitutive response of Mg polycrystals with a pronounced rolling texture, as shown in Fig.~\ref{fig:Polycrystal_Stress}. To this end, a dominant texture in the ND with random rotations around the crystal's $c-$ axis was applied. This behavior emphasizes the anisotropy commonly observed in Mg. While the initial texture significantly impacts the polycrystal's behavior \citep{Zhao2016, Kelley1967, Wonsiewicz1967, Beyerlein2004}, exploring the influence of texture is beyond the scope of this study. Therefore, the same initial texture is selected for each of the simulated polycrystalline samples.

\subsection{Numerical solution for data generation}
Data-generation follows from performing multiple deformation-driven material point simulations of the Taylor model. To this end, deformation is applied in $N$ discrete increments over 100 seconds, starting from an undeformed configuration, in \emph{fixed} time increments $\Delta t = t_l - t_{l-1} =  10^{-3}$, where $t_l, t_{l-1} \in[0, t_\text{max}],~ 0< l \leq N$ are two consecutive, discrete points in time, 
\begin{equation}
    \bm F^{(l)} = \bm F^{(l-1)} + \Delta \bm F, 
\end{equation} where we introduce the superscript notation $^{(l)}$ to denote the value at time $t_l$. The temperature varies and is applied similarly. 

The updates for the internal variables ($\bm\epsilon, \bm\lambda$) are computed explicitly via en Euler-forward integration scheme at each increment, so that 
\begin{equation}
  \epsilon_\alpha^{(l)} = \epsilon_\alpha^{(k)} + |~ \Delta t ~ \dot\gamma_\alpha^{(l)}|,
\end{equation}
and
\begin{equation}
    \lambda_\beta^{(l)} = \lambda_\beta^{(k)} + ~ \Delta t ~ \dot\lambda_\beta^{(l)},
\end{equation}
respectively.
The stress follows from differentiation, so that
\begin{equation}
   \bm{P}^{(l)} = \frac{\partial W_\text{Taylor}^{(l)}(\bm{F}^{(l)},\bm{\epsilon}^{(l)},\bm{\lambda}^{(l)},T^{(l)})}{\partial \bm{F}^{(l)}}.
\end{equation}
The Cauchy stress follows from \eqref{eq:Nanson_stress}.
The exact ways of data generation are described in Section \ref{sec:DataGen}.

% \section{Recurrent neural operator-based surrogate model}
\section{Temperature-aware recurrent neural operator }

We aim to develop an efficient surrogate model for the constitutive response of the Mg, which leverages the data generated via the calibrated and tested CP model described above. This task requires finding an operator capable of learning the map between the deformation history of the material and the present stress of the material, capturing the severe, temperature-dependent plastic anisotropy. 

A general map of this type takes the form\footnote{Note that this map could be formulated even more generally, if the presence of heat sources (e.g., due to plastic deformation) and heat conduction within the material were considered. Yet, we neglect those effects in the present study.}
\begin{equation}
\label{eq:mapfs}
    \psi: \{\bm{F}(\tau), T(\tau) : ~ \tau \in [0,t)\} \mapsto \bm{\sigma}(t), 
\end{equation}
where $\psi$ is a generally non-Markovian, history-dependent map for the temperature-dependent constitutive behavior of a material, given a trajectory in the space spanned by deformation gradient and temperature.

While learning a neural network approximation $\mathcal{F}$ of $\psi$ is generally possible, we prefer to instill more physics into the approximation. Inspired by the nature of plasticity, we decouple the elastic and plastic stress contributions by introducing a physically-motivated decomposition of the Cauchy stress tensor according to
\begin{equation} 
\bm \sigma = \sigma^\text{h}\mathbf{I} + \bm \sigma^\text{dev},
\end{equation}
where $\sigma^\text{h} = \frac{1}{3}\textrm{tr}(\bm \sigma)$ denotes the hydrostatic pressure (associated with the approximately elastic volumetric deformation), whereas $\bm \sigma^\text{dev} = \bm \sigma - \sigma^\text{h} \mathbf{I}$ is the deviatoric stress tensor (associated primarily with plastic shear). $\mathbf{I}$ represents the identity tensor. Motivated by common plasticity models, we hypothesize that the deviatoric stress tensor depends on unknown internal variables, whereas the hydrostatic pressure is typically assumed to have no history dependence, being a function solely of the current temperature $T(t)$ and volume change $J(t) = \det\bm{F}(t)$. However, to ensure a robust model, we adopt a general formulation that permits the hydrostatic stress $\sigma^\text{h}$ to depend on the history of volume change and temperature. This hypothesis will be tested in Section~\ref{sec:Internal variables} to determine whether the no-history assumption holds for the material under study.

The above physically motivated decomposition of the problem allows for a description that maps the deformation and thermal history to the deviatoric stress tensor through  
\begin{equation}\label{eq:map1}
{{\psi}}^\text{dev}: \{\bm{F}(\tau), T(\tau): \tau\in [0,t) \} \mapsto {\bm \sigma^\text{dev}(t)},
\end{equation}
while an independent map $\psi^\text{h}$ maps the volume change $J$ to the scalar hydrostatic stress $
\sigma^\text{h}$:
\begin{equation} \label{eq:map2}
{{\psi}}^\text{h}: \{J(\tau), T(\tau): \tau\in [0,t) \} \mapsto { \sigma^\text{h}(t)}. 
\end{equation}
% Both of these maps are learned independently, facilitating training, as discussed in Section \ref{sec: training}.

% We here implicitly assume that the problem is Markovian, i.e. the future state of the material follows directly from the previous state of the material and that an evolution law exists to predict the future . 

% Given the assumptions made in the homogenization of the material via the Taylor approach, we narrow our attention to a reduced map for the present problem, namely
% \begin{equation}
% \label{eq:mapfs}
% \psi^{\dagger}: \{\bm{F}(t), T(t), \bm{\xi}(t), ~ \bm{x}\in \Omega , ~ t \in [0,t_\text{max}] \}\mapsto \{ \dot{\bm \xi}(t), \bm \sigma(t)\} ,\end{equation} 
% 

% Note that due to the non linearity of $W_i$, $\bm \sigma$ is generally a non linear function of the set of history-dependent internal variables $\bm{\epsilon}, \bm{\lambda}$. 

% where $\sigma = (\textrm{det} F^{-1}SF^T)$ is the Cauchy stress in the macroscale.

% \subsection{Recurrent neural operators}
\subsection{Architecture}

\label{sec:RNO_Def}
To learn the maps \eqref{eq:map1} and \eqref{eq:map2}, we exploit the concept of a recurrent neural operator (RNO), originally proposed in \citet{bhattacharya2022learning,liu2022learning}. Neural operators have numerous advantages compared to other network architectures \citep{kovachki2021neural}. Among others, they provide a discretization-invariant universal approximation of a map between two function spaces. As such, neural operators are ideally suited to learn the above maps.

For our problem, we propose the temperature-aware recurrent neural operator (TRNO), which extends the original (temperature-independent) formulation of \citet{bhattacharya2022learning} to incorporate the thermal history. Specifically, we consider approximations of the form 
\begin{equation} 
\label{eq:rno_theory}
\begin{cases} 
\bm \sigma(t) = \mathcal{F}\big(\bm{F}(t), T(t), \bm{\xi}^\text{dev}(t)\big) +  \mathcal{H}\big(J(t), T(t), \bm{\xi}^\text{h}(t)\big)\bm{I}\\
\dot{\bm \xi}^\text{dev}(t) = \mathcal{G}\big(\bm{F}(t), T(t), \bm \xi^\text{dev}(t)\big), \\
\dot{\bm \xi}^\text{h}(t) = \mathcal{I}\big(J(t), T(t), \bm \xi^\text{h}(t)\big),
\end{cases}
\end{equation}
where $\mathcal{F}: \mathbb{R}^{d\times d} \times \mathbb{R} \times \mathbb{R}^k \mapsto \mathbb{R}^{d \times d}$, $\mathcal{H}: \mathbb{R}^{d\times d} \times \mathbb{R} \times \mathbb{R}^m \mapsto \mathbb{R}$, $\mathcal{G}: \mathbb{R}^{d\times d} \times \mathbb{R} \times \mathbb{R}^k \mapsto \mathbb{R}^{k}$, and $\mathcal{I}: \mathbb{R}^{d\times d} \times \mathbb{R} \times \mathbb{R}^m \mapsto \mathbb{R}^{m}$ are neural networks. We aim to make the problem Markovian by introducing the state vector $\bm{\xi}=\bm{\xi}^\text{dev}\oplus\bm{\xi}^\text{h}\in \mathbb{R}^{k+m}$, which stores the history necessary to describe the state of the material at any time. Hence, internal variables $\xi_i$ aim to encapsulate the thermal and deformation history of the material.
For a rigorous derivation and explanation, we refer to \citet{bhattacharya2022learning}. As done there, we do not impose any thermodynamic or material-symmetry restrictions on the functions in~\eqref{eq:rno_theory}. 
 
We employ fully connected neural networks (FCNNs) for the mappings $\mathcal{F}, \mathcal{H}, \mathcal{G}, \text{and } \mathcal{I}$, which are responsible for learning the stress components and the evolution of the state variables based on the current deformation and $\bm{\xi}$-vector. Given a sequence of input functions $\{\bm{F}^{(j)}, T^{(j)}; j\in \{0,...,N\}\}$ discretized with constant time steps $\Delta t = t^{(j)} - t^{(j-1)}$, we construct the TRNO using a forward-Euler time integration scheme:
\begin{equation} \label{eq:rno}
\begin{cases} 
\bm \sigma^{\text{dev}(j)} = \mathcal{F}(\bm{F}^{(j)}, T^{(j)}, \bm \xi^{\text{dev } (j)}  ) 
% +  \mathcal{H}^\text{FCNN}\big(\det\bm{F}^j, T^j, \bm{\xi}^{\text{h}(j)}\big)\bm{I}
\\
\xi^{\text{dev }(j)}_i = \xi^{\text{dev }(j-1)}_i + \Delta t \ \mathcal{G}_i(\bm{F}^{(j-1)}, T^{(j-1)}, \bm{\xi}^{\text{dev }(j-1)}) 
\quad\text{for } i = 1,\ldots,k.
\end{cases}
\end{equation}
An analogous TRNO is constructed for the hydrostatic components:
\begin{equation} \label{eq:rno}
\begin{cases} 
\sigma^{\text{h}(j)} = \mathcal{H}(J^{(j)}, T^{(j)}, \bm \xi^{\text{h} (j)}  )
\\
\xi^{\text{h}(j)}_i = \xi^{\text{h}(j-1)}_i + \Delta t \ \mathcal{I}_i(\bm{J}^{(j-1)}, T^{(j-1)}, \bm{\xi}^{\text{h }(j-1)}) 
\quad\text{for } i = 1,\ldots,m.
\end{cases}
\end{equation}
Note that, while the equation of state in the physical model is not history-dependent, the above TRNO is required to capture the thermal history.

\subsection{Internal variables and Markovian model}

In the presented architecture, the map $\mathcal{G}$, is charged with the identification of the evolution of (internal) state variables $\bm\xi^\text{dev} \in \mathbb{R}^{k}$, which encapsulate the memory/history of the thermal and deformation path at each time. Analogously, $\mathcal{I}$ is responsible for learning the evolution and identification of $\bm\xi^\text{h}$. In this Markovian formulation of the problem, the state of the material is assumed to be perfectly described by the internal variables. Consequently, at any time $t \in [0,t^\text{max}]$, the evaluation of the RNOs only requires the known inputs $\{\bm{F}(t), T(t), \bm{\xi}(t) \}$ at the current time instead of the complete history. It is important to note that we assume the existence of the internal variables but do not make any assumptions about their number nor their evaluations a priori. Instead, the RNO is trained with different numbers of internal variables (different values of $m$ and $k$) and sought to interpret their physical meaning from the trained networks (and admit the identification of optimal choices of $m$ and $k$). This hence requires a training. 

\subsection{Data generation}
\label{sec:DataGen}
The training and test data were generated from simulations, as explained in Section~\ref{sec:Taylor_CP}. To sample the set of permissible input functions, we adopt the approach described by \citet{LIU2022104668}. A set of random deformation and temperature paths is created, representative of the wide range of possible mechanical behavior. 

To this end, the total time $t_\text{max} = 100 s$ over which deformation will be applied is divided into $N=10^4$ intervals of equal time steps $\Delta t = t_\text{max}/N$. For each path and at each time $t_{j}$ with $j \in [1,N]$, we compute the deformation gradient and temperature, respectively, as
\begin{align} 
\label{eq:Deformationpath}
    F^{(j)}_i &= F^{(j-1)}_i + \nu^{(j)}_i \epsilon_\text{max} \sqrt{\Delta t}, \\
    \label{eq:Thermalpath}
    T^{(j)} &= T^{(j-1)} + \zeta^{(j)} \theta_\text{max} \sqrt{\Delta t},
\end{align}
where $\nu^{(j)}_i, \zeta^{(j)}  \in \{-1,1\}$ follow a Rademacher distribution \citep{LIU2022104668}, $\epsilon_\text{max}>0$ and $\theta_\text{max}>0$ are chosen parameters. At time $t=0$, $\bm F^{(0)} = \bm{I}$, and $T^{(0)} \in [0,250]$ ($^\circ$C). For each pair $(\bm{F}^{(j)},T^{(j)})$ on a given deformation path, the stress tensor is computed as well as its deviatoric and volumetric contributions via the Taylor CP model introduced in Section~\ref{sec:Taylor_CP}. A total of $10{,}038$ individual stress-strain trajectories were generated with $\epsilon_{max} = 0.1$ and $\theta_{max} = 150$. Examples of random strain and temperature paths are given in Fig.~\ref{fig:strain_temp_paths}(a).

\subsection{Loss function}
Let $\bm{\sigma}^{(j)}$ be the ground truth of the stress, as predicted by the CP model, and $\bm{\hat{\sigma}}^{(j)}$ be the corresponding prediction of the RNO at time $t=t_j$. The train and test error are computed as the error over all $N$ time steps and over each of the $N_s$ loading paths with the loss function given by

\begin{equation}
\text{error} = \frac{1}{N_s} \sum_{n_s = 1}^{N_s}\Biggl ( \sum_{j=1}^{N}\frac{|\bm{\sigma}^{(j)}_{n_s} -  \bm{\hat{\sigma}}_{n_s}^{(j)}|^2}{| \bm{\sigma}_{n_s}^{(j)}|^2} \Biggr )^{1/2},
\end{equation}
where $|\cdot|$ denotes the Frobenius norm. 
As the loss function does not include any information about the internal variables $\bm \xi$, in the current architecture the internal variables and their evolution laws are learned implicitly in an unsupervised manner, based solely on the performance of the network to capture constitutive behavior. 

\subsection{Comparison of TRNO, GRU, and LSTM}
\label{sec: training}
\begin{figure}[ht]
    \centering    \includegraphics[width=0.85\textwidth]{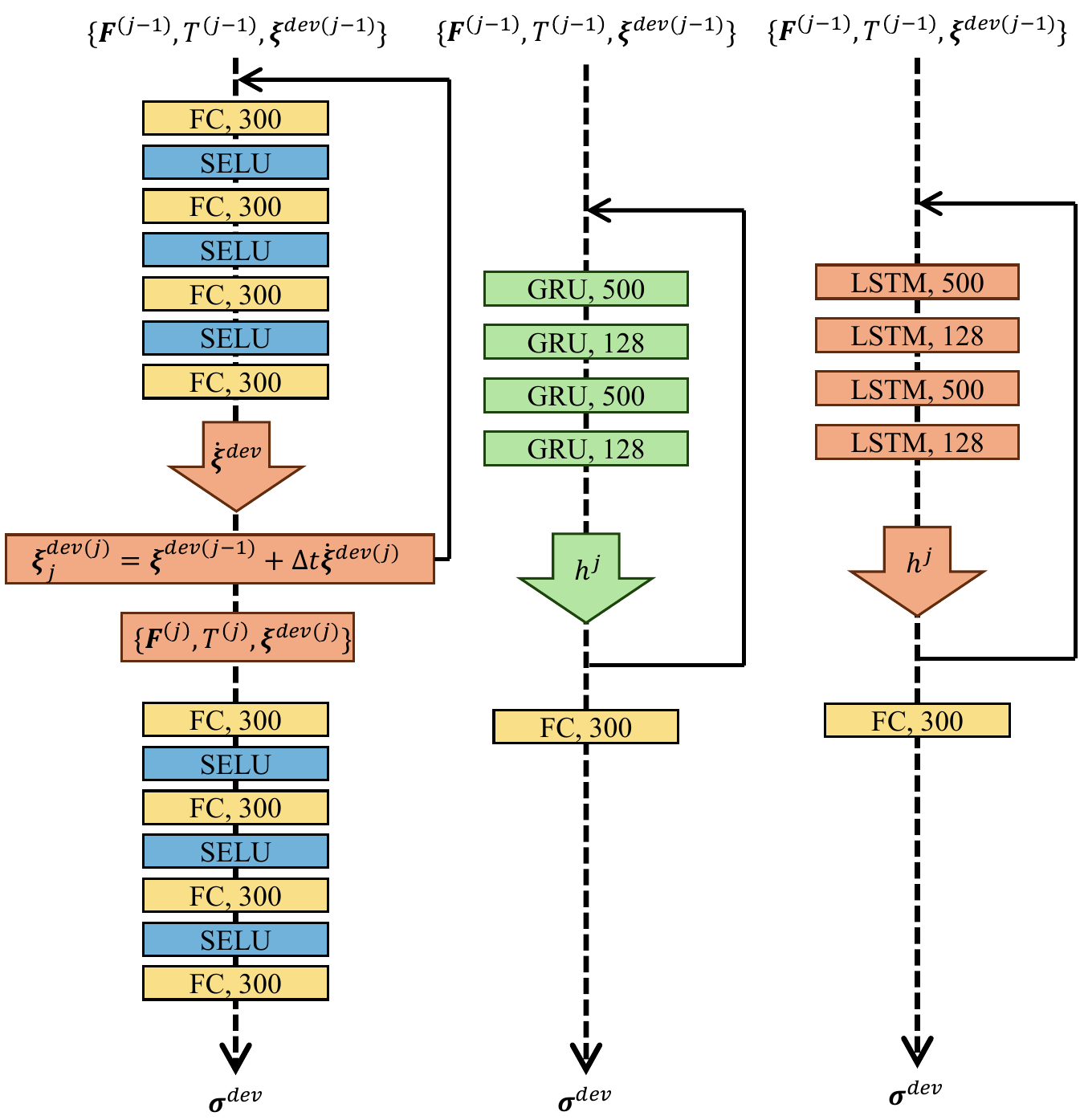}
    \caption{From left to right: Schematic of the TRNO architecture, the GRU, and the LSTM networks for comparison.}
\label{fig:sketch}
\end{figure}

Fig.~\ref{fig:sketch}(a) provides a graphical representation of the TRNO architecture for the deviatoric part (the hydrostatic part is analogous). The networks $\mathcal{F}$ and $\mathcal{G}$ comprise four fully connected layers with Scaled Exponential Linear Units (SELU) activation functions\footnote{Although ReLU is typically the default choice for a hidden layer's activation function, better results could be achieved with SELU. Hyperbolic tangent and sigmoid functions were also tested, but showed limited success.} \citep{klambauer2017self}.
To test the effectiveness of our model, we compare it with classical neural networks based on Gated Recurrent Units (GRU) and Long Short-Term Memory (LSTM) units, which were explored in previous studies for constitutive modeling and plasticity \citep{yu2022elastoplastic, mozaffar2019deep}. For completeness, the details of the GRU and LSTM architectures used in our benchmarks are included in Figs.~\ref{fig:sketch}(b) and (c), respectively. For the TRNO $\psi^\text{dev}$, we use four fully connected layers with 300 nodes each for both networks $\mathcal{G}^\text{dev}$ and $\mathcal{F}^\text{dev}$. The same architecture is used for the TRNO $\psi^\text{h}$. All training was completed using Adam Optimizer with the Cosine Annealing Scheduler in PyTorch \citep{paszke2017automatic, kingma2017adam}.

\section{Results}

We demonstrate the performance of the TRNO presented in Section~\ref{sec:RNO_Def} by applying it to the temperature-dependent response of polycrystalline Mg described by the model of Section~\ref{sec:Background_CP}. A total of $10{,}038$ samples were generated, using the random input functions defined in \eqref{eq:Deformationpath}, from which we use $1{,}538$ for testing and the remaining $8{,}500$ for training.

\begin{figure}[t]
    \centering    \includegraphics[width=0.9\textwidth]{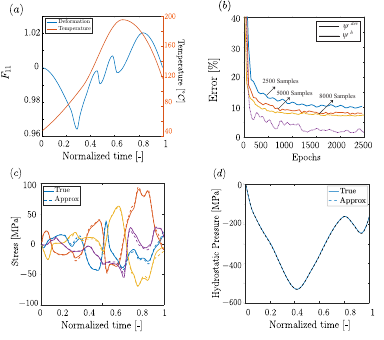}
    \caption{(a) Typical temperature and strain loading paths for data generation. (b) Test errors for the TRNO $\psi^{dev}$ trained with varying data sizes, as well as the test error for the TRNO $\psi^h$ trained with 8000 samples. (c) True evolution (solid lines) and TRNO approximation (dashed lines) of the deviatoric stress of a compressed sample. (d) True hydrostatic pressure and its TRNO prediction. }
    \label{fig:strain_temp_paths}
\end{figure}

Representative results are shown in Fig.~\ref{fig:strain_temp_paths}. Typical input functions $\{ \bfF(\tau), T(\tau) \}_{0 \le \tau \le t}$ are presented in Fig.~\ref{fig:strain_temp_paths}(a). Note that the figure shows only $F_{11}$ for simplicity, while the remaining components of the deformation gradient follow the same random evolution. The TRNO test errors of $\psi^\text{dev}$ and $\psi^\text{h}$ for the loading path in Fig.~\ref{fig:strain_temp_paths}(a) are shown in Fig.~\ref{fig:strain_temp_paths}(b) for selected sizes of the training dataset. The TRNO prediction of $\sigma^\text{dev}$ and $\sigma^\text{h}$ versus the normalized time $\bar{\tau} = \frac{\tau}{t}$ is shown in Fig.~\ref{fig:strain_temp_paths}(c) and (d), respectively, and compared to the true (simulated) response. Results were obtained with a fixed internal variable dimensions $m = 35 $ for the deviatoric part, and $l = 0$ for the hydrostatic part. The identification of internal variables will be discussed in detail in Section~\ref{sec:Internal variables}. 
We observe that with $8{,}000$ training samples, the test error for $\psi^\text{dev}$ is about 6.2\%, while the test error for $\psi^\text{h}$ is approximately 2.9\%, thus confirming good accuracy of the TRNOs to predict the response of the temperature-dependent CP model during the combined thermo-mechanical loading. 

\subsection{Generalization by transfer learning}

A key concern of using ML-based methods is the ability of the trained models to generalize to unseen functions, here to unseen thermo-mechanical loading paths. To this end, we seek to further validate the trained TRNO with input functions generated from different probability measures than those defined in \eqref{eq:Deformationpath} and \eqref{eq:Thermalpath}. We first construct a test dataset of $9{,}965$ samples with a different thermal history, in which the temperature is kept constant throughout the time interval of interest\footnote{This dataset is generated by setting $\zeta_j = 0  \ \forall \ j$.}, so
\begin{equation}
T(\tau) = T_c \in \mathbb{R}, \quad 0\leq \tau \leq t .
\end{equation} 

\begin{figure}[!b]
    \centering
    \includegraphics[width=0.9\textwidth]{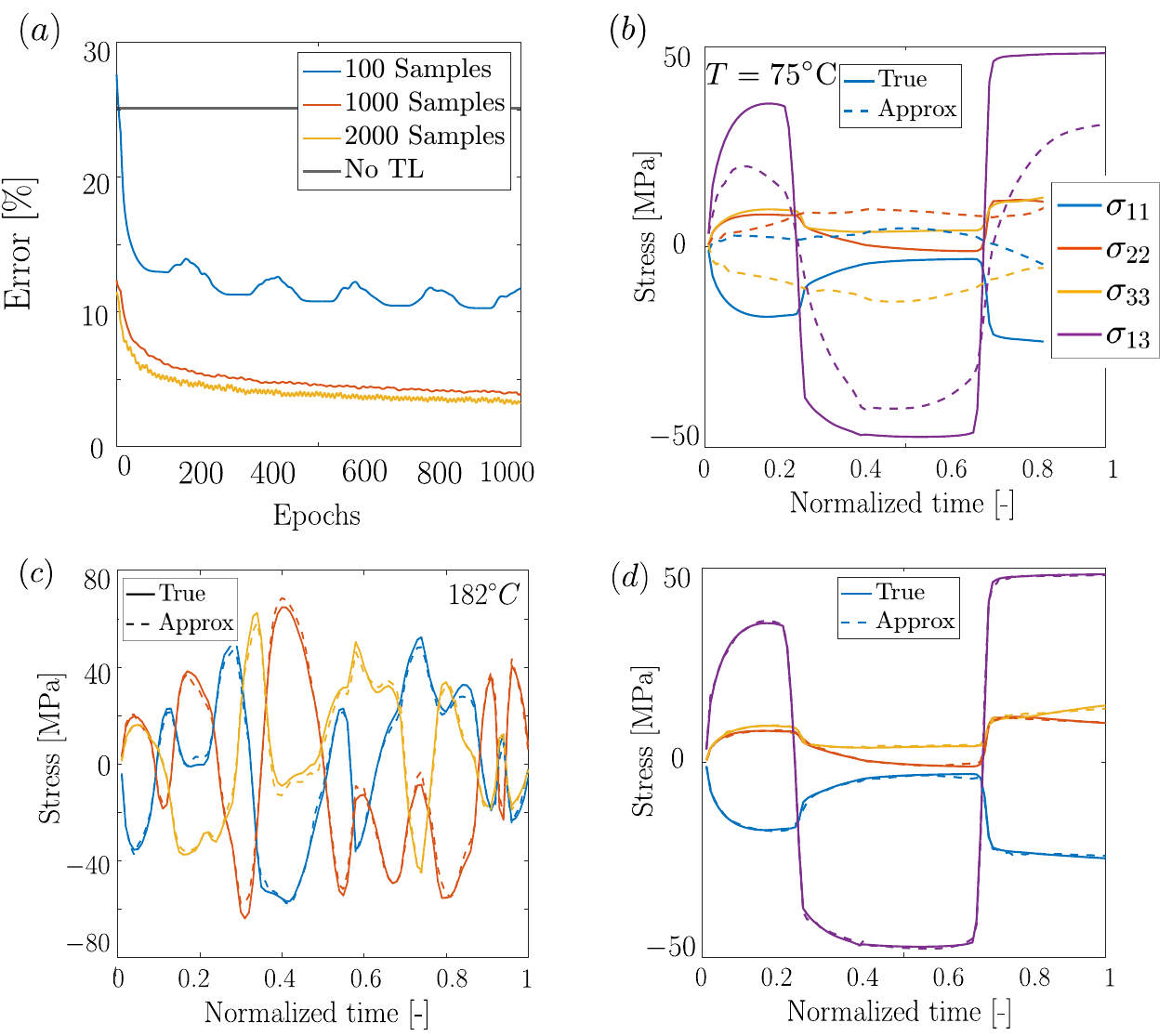}
    \caption{(a) Error during transfer learning (TL) for multiple samples. (b) Ground truth (solid lines) and RNO prediction (dashed lines) before transfer learning and (c,d) and after transfer learning for various temperatures and different randomized and semi-deterministic loading paths.}  
    \label{fig:TrainingErrors_TC}
\end{figure}

The trained TRNO is tested with this dataset, the results of which are shown in Fig.~\ref{fig:TrainingErrors_TC}(a,b). Strikingly, the TRNO trained with a random thermal history performs poorly when tested on a constant-temperature path. As a remedy, we use transfer learning (TL) \citep{bozinovski1976influence,torrey2010transfer} and fine-tune the trained TRNO, using the constant-temperature dataset. The test error before and after transfer learning is included in {Fig. \ref{fig:TrainingErrors_TC}}(a). We observe that with a relatively small amount of data (100 samples) transfer learning reduces the test error from 25\% to 10\%, showing the efficacy of the transfer learning method. Increasing further to $1{,}000$ training samples leads to a better prediction of material behavior. Beyond that, the addition of more training data only have a marginal effect, as evidenced for the case of 2,000 training samples. 

Although TRNO trained with random thermal paths generalizes poorly to other thermal paths, the fine-tuned TRNO after transfer learning is capable of generalizing to other deformation paths, including those that are more commonly encountered in practice. For example, Fig.~\ref{fig:TrainingErrors_TI} summarizes results for uniaxial loading, showing convincing agreement between the RNO predictions and the true (simulated) responses. 

\subsection{Anisotropy and temperature dependence}

\begin{figure}[h]
    \centering
    \includegraphics[width=0.9\textwidth]{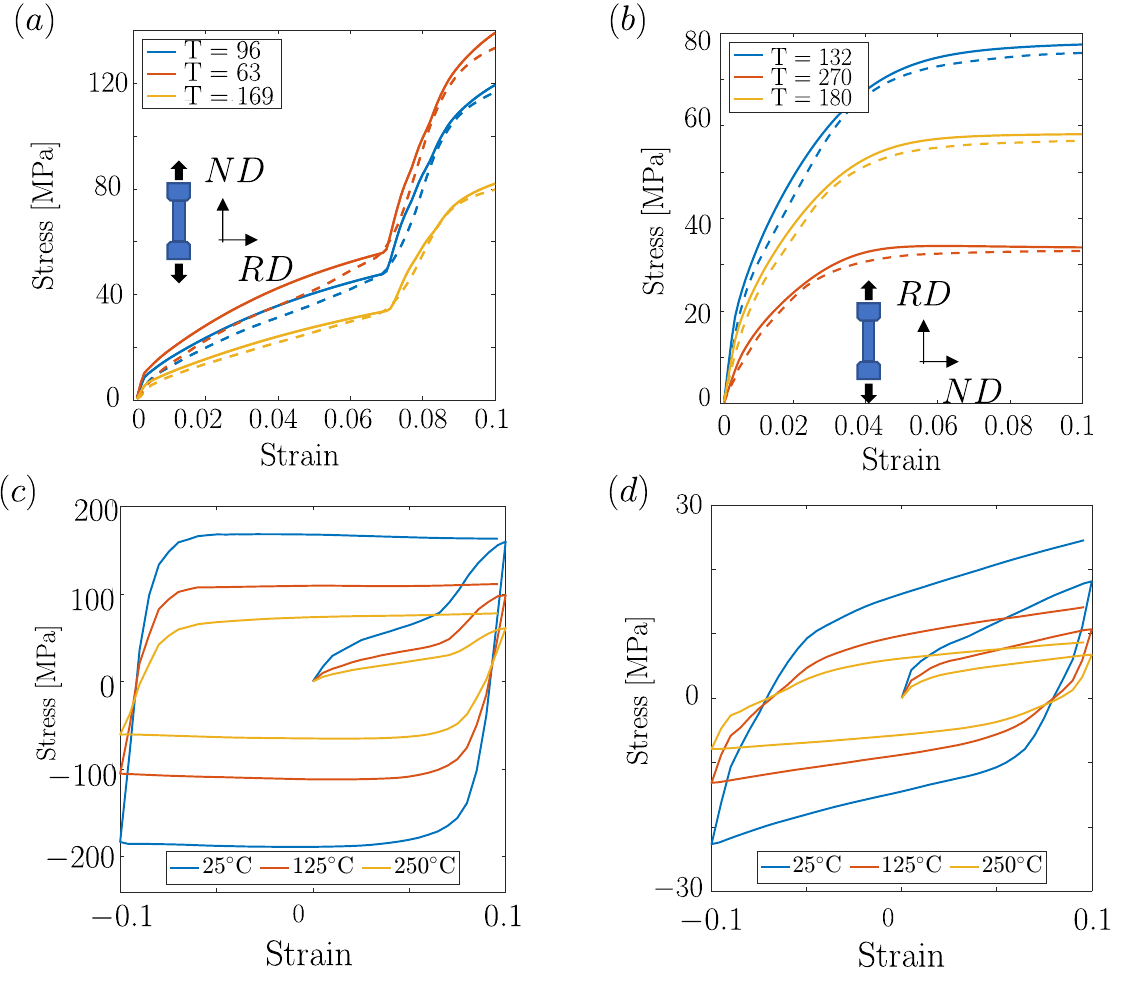}
    \caption{Ground truth (solid lines) and RNO predictions (dashed lines) of the material behavior for multiple temperature levels in (a) tension along the ND and (b) tension along the RD. (c) and (d) RNO-predicted responses of samples cyclically loaded in tension along the ND and shear along the RD, respectively.}  
    \label{fig:TrainingErrors_TI}
\end{figure}

Recall from Section~\ref{sec:material behavior} that the material's plastic response is both anisotropic and temperature-dependent (cf.~Fig.~\ref{fig:Polycrystal_Stress}). We evaluate the TRNO's capability to capture those effects in Fig.~\ref{fig:TrainingErrors_TI}, by investigating three different temperature levels and two loading directions. The polycrystals in Fig.~\ref{fig:TrainingErrors_TI}(a) were loaded along the ND of the textured sample (i.e., in the direction with which most grains' $c-$~axes align; cf.~Fig.~\ref{fig:Polycrystal_Stress}(c)). The response exhibits the typical twin-dominated behavior with a sharp change at abound 7\% strain. In contrast, Fig.~\ref{fig:TrainingErrors_TI} (b) depicts the stress-strain response of polycrystals loaded in the RD, thus not exhibiting pronounced twinning. The TRNO captures the results accurately, both qualitatively and quantitatively. Fig.~\ref{fig:TrainingErrors_TI}(c) shows the results of a polycrystal under cyclic loading along the ND at three distinct temperatures. We observe that the TRNO captures the Bauschinger effect and the temperature-dependent softening of the material. The initial tensile load cycle shows the twin-dominated response, followed by a change to compressive stress levels, as we may expect. In Fig.~\ref{fig:TrainingErrors_TI} (d), the analogous results are presented for shear along in the $12-$direction, showcasing the versatility of our TRNO.

\subsection{Internal variables}
\label{sec:Internal variables}

\begin{figure}[!h]
    \centering
    \includegraphics[width=\textwidth]{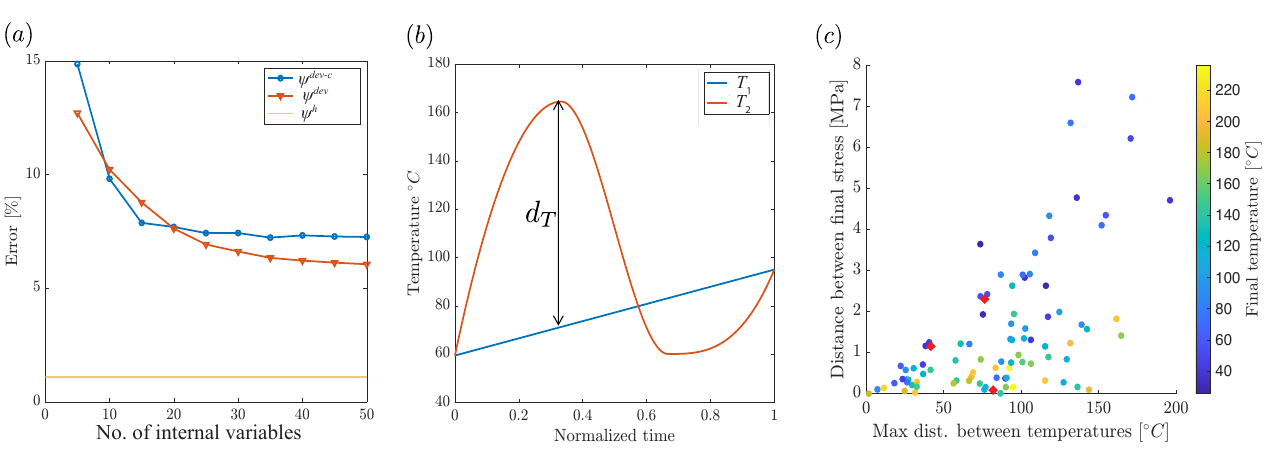}
    \caption{(a) Test error of the TRNO predictions vs.\ number of internal variables, for TRNOs $\psi^{dev}$ and $\psi^h$, as well as that for TRNO $\psi^{dev-c}$, which is trained with a data set with varying deformation paths and fixed temperature paths. (b) Illustration of a pair of thermal paths $T_1$ and $T_2$ with the same initial and final temperatures. (c) Distribution of the stress difference (in MPa) between the linear vs.\ varying temperature paths when subjected to the same deformation paths for each sample. Here, 100 pairs of randomized temperature histories are shown. Circular dots represent TRNO predictions, whereas red diamonds show the results from the CP model for comparison.}
    \label{fig:IV_Evolution_TI}
\end{figure}

While the internal variables in the CP model have physical meaning, those in the TRNO generally do not. Therefore, one question we seek to answer in this study is whether the TRNO can be equipped with a finite set of internal variables capable of fully characterizing the thermal and deformation history of the polycrystalline crystal plasticity model. To provide an answer, we train a family of TRNOs with varying dimensions $k$ and $m$ of the internal variables $\bfxi^\text{dev}$ and $\bfxi^\text{h}$, respectively, in \eqref{eq:rno}. The results are summarized in Fig.~\ref{fig:IV_Evolution_TI}(a), which reveals interesting observations. The training of $\psi^h$ exhibits a constant test error irrespective of the number of internal variables. This suggests that the macroscopic hydrostatic stress does not depend on deformation or thermal history, a finding consistent with conventional assumptions in plasticity theory.

By contrast, there is a clear dependence of the deviatoric stress $\sigma^\text{dev}$ on the deformation and thermal history, as evidenced in Fig.~\ref{fig:IV_Evolution_TI}(a), where the test error decreases significantly with an increasing number of internal variables, until reaching a plateau when the number of internal variables becomes large.

We further investigate whether our model is capable of learning the thermal history dependence from data. To this end, we consider the stress-strain response of the material along a fixed deformation path but with different thermal paths, fixing only the initial and final temperature values. Let $T_1, T_2: [0, t] \mapsto \mathbb{R}$ be a pair of thermal paths, constrained by $T_1(0) = T_2(0)$ and $T_1(t) = T_2(t)$, where $T_1$ is a random path defined in Section~\ref{sec:DataGen}, while $T_2$ is a linearly varying path connecting the start and end points of $T_1$ (see Fig.~\ref{fig:IV_Evolution_TI}(b)). We measure the distance between the pair of paths by the $L_{\infty}$-norm:
\begin{equation}
d_T = \|T_1 - T_2\|_\infty = \sup_{\tau \in [0,t)} |T_1(\tau) - T_2(\tau)|.
\end{equation}
We define the distance $d_\sigma$ between the corresponding deviatoric stress states as the absolute difference at the final time $t$ as
\begin{equation}
d_\sigma = |\bm{\sigma}_1(t) - \bm{\sigma}_2(t)|.
\end{equation}
The TRNO's prediction of $d_{\sigma}$ for a total of 100 pairs of randomly generated thermal paths versus increasing values of the maximum temperature difference $d_T$ is shown in Fig.~\ref{fig:IV_Evolution_TI}(c), along with selected predictions from the Taylor CP model (shown in red). We observe that the TRNO is capable of learning the path dependency, evidenced by the fact that $d_{\sigma} \neq 0$ for almost all of the pairs tested. Additionally, there is a positive correlation between $d_{\sigma}$ and $d_T$, especially when the final temperature is low.

To further investigate whether additional internal variables are required to account for the thermal history, we trained a TRNO $\psi^{dev-c}$ with exactly the same architecture as $\psi^{dev}$, but using a dataset whose temperature paths are constant in time while keeping the deformation paths random, as defined in \eqref{eq:Deformationpath}. The corresponding results for $\psi^{dev-c}$, trained with varying numbers of internal variables, are shown in Fig.~\ref{fig:IV_Evolution_TI}(a). Intriguingly, we observe an earlier and sharper transition to the plateau regime above 15 internal variables for the TRNO trained with temperature-independent data, which suggests that additional internal variables are helpful for the TRNO to better approximate the temperature-varying data. 

% However, we do not observe a significant decrease in the test error with increased internal variables above 15, which suggests that the present set of internal variables suffices to characterize the constitutive response of the material.

\subsection{Benchmarks and resolution independence}
\label{sec:Resolution Indep}
\begin{figure}[h!]
    \centering
    \includegraphics[width=\textwidth]{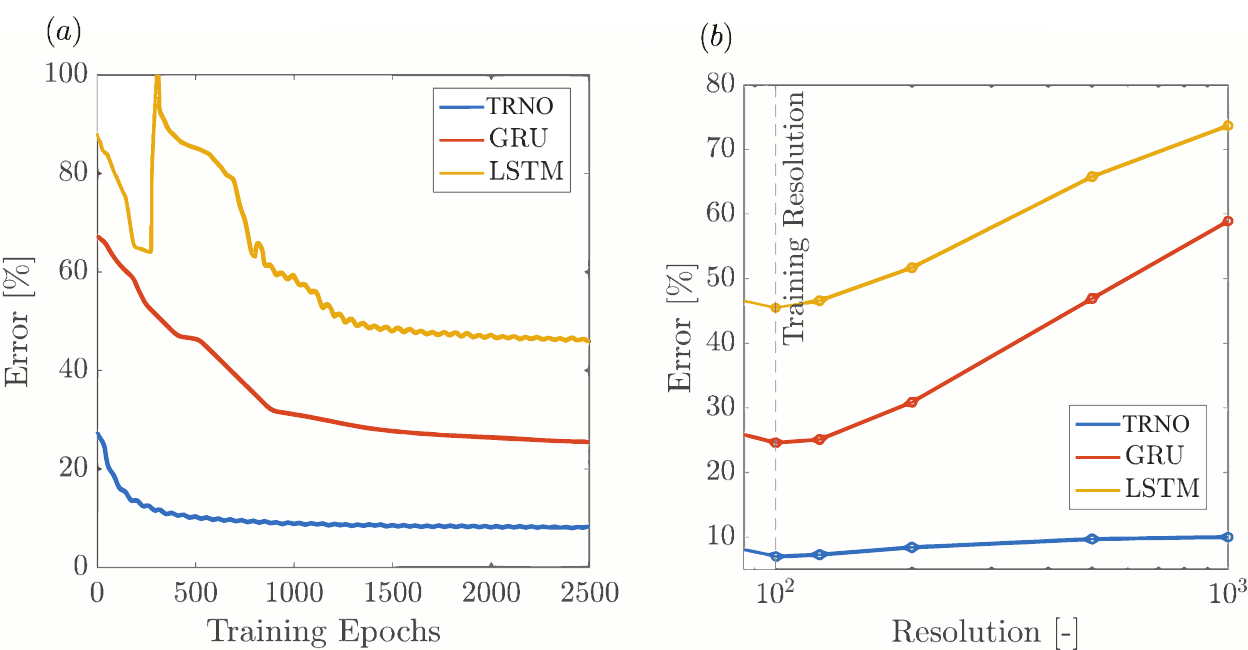}
    \caption{Performance and resolution independence. (a) Test error versus training epochs for TRNO, GRU and LSTM. (b)~Comparison of test error of trained TRNO, GRU and LSTM at a resolution of 200 time-steps against test data of various resolutions. }
    \label{fig:Resolution}
\end{figure}

The efficacy of the TRNO architecture is tested against benchmark ML models based on GRUs and LSTM from the literature \citep{yu2022elastoplastic,mozaffar2019deep}. A schematic of the LSTM and GRU architectures is shown in Fig.~\ref{fig:sketch} along with the RNO. 
The three architectures are tested against the ground truth of the CP model with the same $8{,}500$ training data. The test error versus the number of training epochs for the three architectures is shown in Fig.~\ref{fig:Resolution}(a). For fairness, the total number of learning parameters is fixed to 8500 for all three models. Remarkably, the TRNO outperforms the other two models significantly. In addition, we notice that the number of hidden variables required by the LSTM and GRU to achieve the minimum test error of Fig.~\ref{fig:Resolution}(a) is 128 \citep{yu2022elastoplastic}, while the TRNO required only 15 internal variables for the same level of accuracy.

The primary advantage of the TRNO architecture is its capacity for zero-shot super-resolution. That is, once trained with input-output pairs from a given temporal  resolution (i.e., time step size), TRNOs are capable of generalizing to input/outputs with different resolutions than those of the training data, without the need of retraining or transfer learning. We examine this feature by testing the trained TRNO $\psi_r$ at a resolution of 100 time steps against a variety of different input resolutions generated from fixed random loading paths, as defined in~(\ref{eq:Deformationpath}) with $t_{max} = 100 \text{s}$. The results are shown in Fig.~\ref{fig:Resolution}(b). The RNO is largely independent of the data resolution, therefore providing a resolution-\textit{independent} approximation of the mapping $\psi$. By contrast, Fig.~\ref{fig:Resolution}(b) shows that the LSTM- and GRU-based models perform more poorly in generalizing to other resolutions, as evidenced by the steep increase in testing errors when moving away from the training resolution. The fact that RNOs are independent of the data resolution (i.e., they enable freely choosing the macroscopic time step without being constrained by a fixed training resolution) makes them well suited for multiscale simulations, which will be demonstrated in Section~\ref{sec:Abaqus_Dynamic}.

\section{Multiscale modeling}
\label{sec:Abaqus_Dynamic}

A major benefit of using TRNO-based surrogate material models lies in the enormous time savings in evaluating otherwise complex material models, which makes them ideally suited for large finite element and multiscale simulations. Physics-based models (such as the Taylor CP model used in this work) are often based on many internal variables, whose updates are time-consuming. Especially in dynamic simulations, where explicit time-stepping schemes may require small time steps, those models quickly become prohibitively expensive when used at the material-point level of finite element simulations. The TRNO presented here was trained efficiently based on constant-time-step data, but (thanks to the properties discussed in Section~\ref{sec:Resolution Indep}) can be used in complex simulations with adaptive time stepping. 

To showcase the performance and potential applications of our TRNO architecture, we present in the following examples of finite element simulations, for which we implemented the trained TRNOs with transfer learning as a material model ("VUMAT") in the commercial finite element package ABAQUS \cite{abaqus}. We highlight that these calculations would not have been possible with the underlying Taylor CP models in reasonable computing times. Because the local material model is derived from a Taylor-type polycrystal (which represents the material behavior on the microscale), one may regard this as a multiscale problem. Where possible, we will give estimates of the runtime improvements due to the use of the TRNO as a surrogate model.

\subsection{Taylor anvil test}

The Taylor anvil test is a well-studied benchmark, which simulates the impact of a cylindrical rod against a rigid wall. We consider cylinders of diameter $D = 1$~mm and height $H = 5$~mm, which are taken from a rolled Mg sample (having the same texture as before) in the RD and ND directions, such that the majority of the Mg grains are oriented along, respectively, the logitudinal and radial directions (see \label{fig:Taylor_Anvil}(e)). The cylinders are simulated to hit a frictionless, rigid wall at time $t = 0$ at an incoming velocity $V = 200\, \text{ms}^{-1}$. We perform the simulations with the Abaqus/Explicit solver, using the TRNO-based material model at the quadrature-point level, at the three temperatures of 25$^\circ \text{C}$, 115$^\circ \text{C}$, and 205 $^\circ \text{C}$ (assuming a constant, uniform temperature) for both orientations of the cylindrical sample. All simulations are conducted using three-dimensional eight-node linear elements with reduced integration (C3D8R in the ABAQUS notation). 

\begin{figure}[h]
    \centering
       \includegraphics[width=0.85\textwidth]{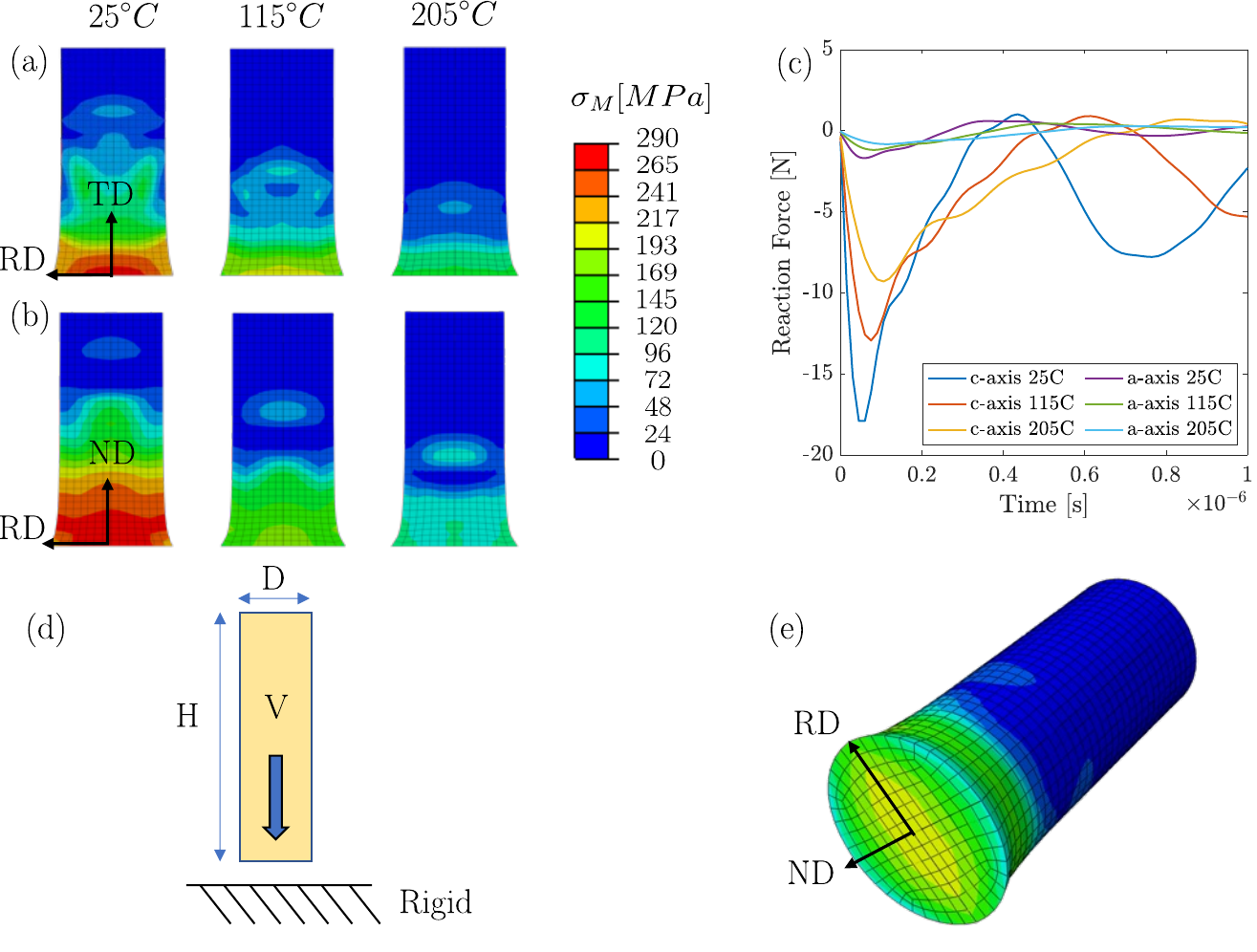}
    \caption{Results of Taylor anvil simulations: (a) von Mises stress distributions in the Mg anvil $10^{-6}$~ms after impact (a) along the TD direction and (b) along the ND direction, each at the three indicated temperature levels. (c) Temperature-dependent evolution of the net reaction force on the impacted face. (d) Sketch of the anvil and experimental setup. (e) The von Mises stress distribution in the sample (same color legend as in (a)), as viewed from the bottom, demonstrates the anisotropy in the sample during impact along the TD at 115$^\circ C$.}
    \label{fig:Taylor_Anvil}
\end{figure}

An elastic wave is transmitted upon impact, followed by plastic deformation. Figures \ref{fig:Taylor_Anvil}(a,b) show cross-sections of the impactor after $10^{-6}$~ms, which illustrate the influence of the sample orientation. 
 
In Figure \ref{fig:Taylor_Anvil}(a), the sample is oriented such that the ND lies in the plane of impact, resulting in compression along the TD, where the $a$-axis poles are predominantly oriented. The deformation is borne by a mixture of basal and prismatic slip, leading to a relatively weak material response. The maximum stress is concentrated at the center of the impact zone, where a combination of compression along the longitudinal direction and tension in the radial direction is experienced. Tension along the RD or TD directions is relatively stiff \citep{jain2007modeling}, compared to the tensile deformation along the ND of the rolled sample, which leads to the asymmetric response we observe in this case. Upon further deformation, twinning would occur and the stress level would increase drastically along the ND. Fig.~\ref{fig:Taylor_Anvil} (e) shows the von Mises stress distribution at $115^\circ$C, revealing that the anisotropy persists at elevated temperatures. 

In contrast, impact along the ND leads to a different overall response, as the material is oriented such that the ND aligns with the impact direction. This leads to a significantly stiffer material response than in the previous case. We observe that the material experiences overall higher stresses in a larger region near the impact zone. The overall stress level decreases rapidly with temperature, which is a well-known phenomenon in Mg \citep{Ono2004}, rendering the overall response more isotropic at elevated temperature. 

Fig.~\ref{fig:Taylor_Anvil}(c) shows the net reaction force (measured across all nodes of the impact face) for all 6 cases. As expected, we observe a large difference between the two orientations as well as the three temperatures. As in the CP model formulation (see Section~\ref{sec:model_singleCP}), the elastic response is also temperature dependent (due to the temperature-dependent elastic moduli), The TRNO model captures this behavior accurately, which becomes evident especially for the ND impact cases. The reaction force rapidly increases, followed by an elastic wave that propagates and reflects at the free end of the sample. The wavelength of the elastic wave changes with temperature, as the bulk modulus decreases with temperature. This feature is captured well by the TRNO $\psi^{h}$.

\subsection{Plate impact test}

As a final application of the TRNO, we simulate an impact test on a plate of Mg along the RD at room temperature, using the commercial software Abaqus and the same finite element setup as in the previous test case. The projectile is assumed rigid and frictionless.
During such plate impact loading, the temperature in the material is known to rise \citep{RAVINDRAN2020101044}, which is in general attributed to two phenomena. First, a portion of the plastic work is converted into heat \citep{ghosh2017plastic,RAVINDRAN2020101044}. A linear relationship is typically assumed \citep{ghosh2017plastic} between the increase in temperature and the dissipated energy, with a proportionality constant known as the Taylor-Quinney factor. For Mg and AZ31B this factor was shown to depend on the orientation of the material \citep{ghosh2017plastic}, and an impact along the RD leads to a low percentage of plastic energy dissipated into heat. Therefore, we ignore this contribution to heating here. The second mechanism leading to material heating is linked to the quasi-isentropic elastic compression of the material \citep{RAVINDRAN2020101044}, which can be described by the incremental relation \citep{grunschel2009pressure} 
\begin{equation}
\label{eq:isentropic}
    \dd T = \frac{a_\text{TH}\exp(-mp) T }{C_p\rho_0}\dd p,
\end{equation}
where $T$ represents the temperature, $a_\text{TH} = 7 \times 10^{-5} \text{K}^{-1}$ and $m = 0.0021 \text{GPa}^{-1}$ are constants, $C_p= 
 (900 + 0.446T) ~ \text{J kg}^{-1} \text{K}^{-1}$ is the specific heat, $p$ is the hydrostatic pressure, and $\rho_0$ is the initial mass density of the material.
We solve \eqref{eq:isentropic} explicitly at every time step of the finite element simulation at each local material point. 

\begin{figure}[h!]
    \centering
     \includegraphics[width=0.85\textwidth]{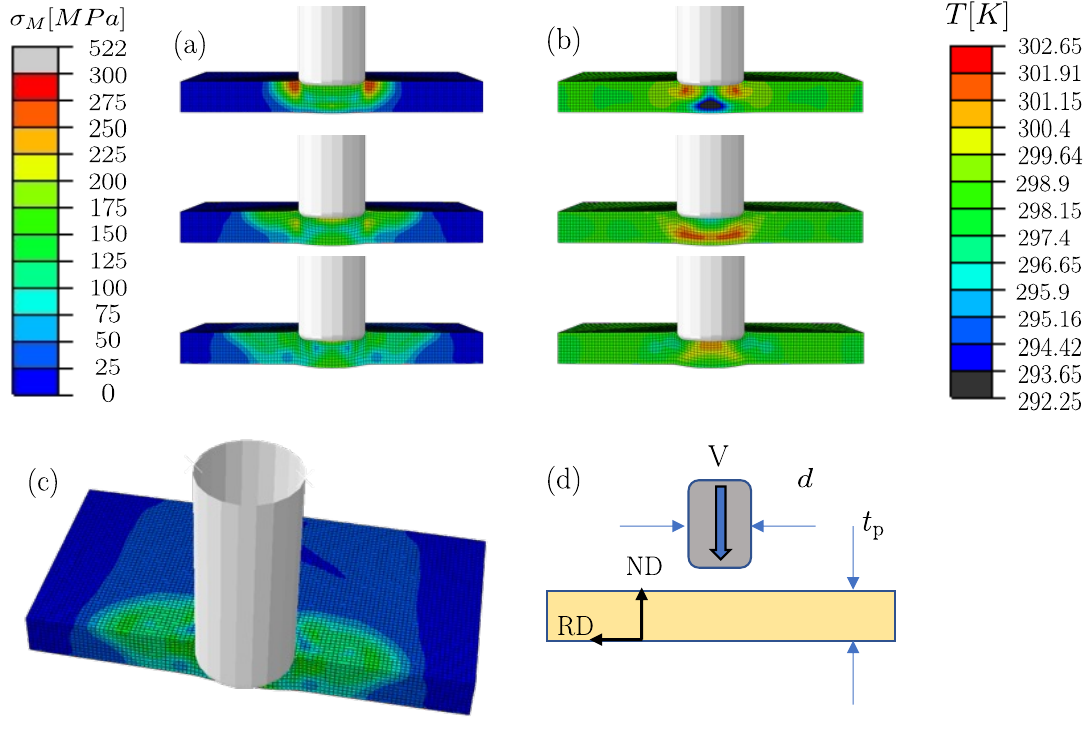}
    \caption{(a) Von Mises stress distribution and  (b) temperature distribution in the simulated Mg plate at (from top to bottom) $10\cdot 10^{-8}, 8\cdot 10^{-8}$, and $1.6\cdot 10^{-7}$~s after impact. (c) Distribution of the von Mises stress, showing the anisotropy in Mg at 25$^\circ C$ during impact. (d) Geometry of the plate impact test.}
    \label{fig:Plate Impact}
\end{figure}

Fig.~\ref{fig:Plate Impact}(a) shows the von Mises stress distribution inside the plate at $10\cdot 10^{-8}, 8\cdot 10^{-8}$, and $1.6\cdot 10^{-7}$~s after impact.
Stress peaks are observed near the edge of the rounded impactor, which relatively quickly dissipate and spread into the material. Fig.~\ref{fig:Plate Impact}(c) highlights the anisotropic stress distribution inside the material. These results compare well with experimental results for pure Mg shown in \cite{RAVINDRAN2020101044}. 
% The isentropic component \citep{RAVINDRAN2020101044} is lower for the RNO-based model due to the significantly lower hydrostatic stresses experienced, limited to roughly 1-2~GPa (Fig.~\ref{fig:Isentropic}).

% \begin{figure}
%     \centering
%     \includegraphics[width=0.5\textwidth]{Temperature_Isentropic.pdf}
%     \caption{Caption}
%     \label{fig:Isentropic}
% \end{figure}

\subsection{Runtime comparison}

To quantify the computational cost savings, we performed a runtime comparison to demonstrate the speedup potential of the TRNO-based surrogate material model relative to the underlying crystal plasticity (CP) model.
To this end, material point computations were performed using the Taylor CP model with 200 grains, as well as the TRNO. Each computation was repeated five times. We first conducted simulations up to 10\% strain with random loading paths defined in~(\ref{eq:Deformationpath}) over 100,000 steps, at a strain rate of $\dot{\epsilon} = 10^{-4}$. The total simulated time was thus 100~s, with a time step of $\Delta t = 10^{-3}$~s. For comparison, we also performed CP simulations with an implicit solver, which enabled significantly larger time steps (and hence fewer computations). Simulations for the TRNO were performed on a single core of a MacBook Pro (2019) with a 2.6~GHz 6-Core Intel Core i7 processor and 16~GB of RAM, whereas the Taylor CP model was run on ETH's high-performance computing cluster, Euler. Both simulations were executed in the same virtual environment on their respective systems. The Taylor model is a highly efficient C++ application, while the TRNO is a non-parallelized PyTorch implementation without CUDA support.
\begin{table}[!t]
    \centering
    \begin{tabular}{c|c|c|c}
         Simulation Type & \# Steps & Wall-clock time $[\text{s}]$ & Threads / Cores\\
         \hline
         Taylor CP model, 200 grains, explicit & $10^5$ & 77.07 & 1 / 1\\ 
         Taylor CP model, 200 grains, explicit & $10^5$ & 8.520 & 16 / 16  \\ 
         Taylor CP model, 200 grains, implicit & $4000$ & 48.7 & 1 \\ 
         TRNO-based model & 100 & 0.0583 & 1 
    \end{tabular}
    \caption{Runtimes of the physical model vs the TRNO for multiple aplications}
    \label{tab:runtime}
\end{table}

Table~\ref{tab:runtime} shows the execution times for the TRNO and Taylor models in single and parallelized applications. Compared to the single-core executable Taylor CP model, the TRNO offers a speedup of $1321.9$; compared to the parallelized 16-core Taylor CP simulation it still represents a speedup of 146 (due to the improved stability and the considerably larger time steps that the RNO can accommodate). Even when enhancing the CP implementation with an implicit solver, the speed up is still a factor of 835. 
In addition, the implicit solver has its own drawbacks, especially in dynamic simulations, and it faces stability issues, as described in \citet{Chang2015}. The TRNO is considerably more robust than the CP models investigated here, which was especially noticeable in the multiscale Abaqus simulations of the Taylor anvil and plate impact tests, where, even with very coarse meshes and severe deformations, the TRNO still converged to a solution, where traditional models would likely fail. This is also the reason why the above Taylor anvil and plate impact examples were presented without results from a CP model for comparison: the computational costs of using the Taylor CP model with a large number of grains at every quadrature point in a finite element simulation would be too expensive (given our limited computational resources) and, moreover, suffer from instabilities and associated issues due to the adaptive (or otherwise prohibitively small) time step used during simulations.

\section{Conclusion}

We have presented a temperature-aware recurrent neural operator (TRNO) as an efficient surrogate for complex material models, which we demonstrated for the example of a polycrystal plasticity model for pure Mg. The TRNO learns incremental updates of the hydrostatic and deviatoric stresses under complex thermo-mechanical loading. Since the TRNO architecture is  Markovian (as the history dependencies are characterized by internal variables that evolve in time), the stress updates can be evaluated locally in space and time. 
We compared the training time and performance of the TRNO with state-of-the-art architectures such as the GRU and LSTM cells and noted that the TRNO achieves a better and more importantly, a time-discretization independent approximation of the constitutive behavior. This is beneficial, among others, for using the TRNO-based surrogate model in large-scale finite element simulations, as shown for multiscale benchmarks that solve a finite element problem on the macroscale while using the polycrstyalline material model at the material-point level.
The TRNO represents the current internal state of the material by a limited set of state variables, which were learned without a-priori knowledge of the physics and whose number was shown to ideally be of the same order as for the physical model. For non-isothermal load paths, additional state variables are required, which indicates a thermal path dependence of the material -- though this requires further investigation. Key conclusions of our benchmark examples are summarized as follows: 
 \begin{itemize}
     \item The TRNO is easier to train and performs more accurately than LSTM- and GRU-based RNNs for the same training data. 
     
     \item The TRNO provides a time-resolution-independent representation of the constitutive behavior, allowing predictions of the material response at resolutions finer than the training resolution, where LSTM and GRU architectures show significantly larger errors. 
     
     \item Under isothermal conditions, the number of state variables required by the TRNO is of the same order as the number of internal variables of the physical model used to generate the training data. Additional state variables are required when temperature varies along the load path, which indicates a temperature-path dependence of the constitutive response.
     
     \item The TRNO performs well on chosen validation data, capturing the overall response accuractely and capturing the salient features of the material's stress-strain response such as the reorientation and stiffenung due to twinning and the anisotropy of the material across the range of temperatures investigated. It also captures the Bauschinger effect across a wide range of temperature and for different load cases. 
     
     \item Integrating the TRNO-based surrogate model in Abaqus/Explicit enabled its use in finite element simulations a significant speedup and improved simulation stability. 

 \end{itemize}

Of course, the present framework has limitations, which offer opportunities for the future. Texture evolution has not been considered but would be an interesting next step, especially for Mg. Similarly, predicting the stress-response for different initial textures was outside the scope of this study. Furthermore, the TRNO as a pre-trained network can easily be applied to other material models or generalized via transfer learning. We also see potential for this architecture in inverse design problems, owing to the speed and efficiency of the TRNO-based surrogate model.

\section*{Declaration of Competing Interests}
None. 
 
 \section*{Acknowledgments}
YH and DMK acknowledge the support from the ETH+ project SynMatLab. BGL gratefully acknowledges the support of Granta Design through the startup fund.
\bibliography{ML_Bib.bib}

% \section*{Appendix}

\end{document}